\documentclass[pra,preprint,superscriptaddress,floatfix,showpacs]{revtex4-1}

\usepackage{amsmath,amsfonts,amssymb}
\usepackage{graphicx,color}

\def\beq{\begin{equation}}
\def\eeq{\end{equation}}
\def\beqn{\begin{eqnarray}}
\def\eeqn{\end{eqnarray}}

\def\bl {\mbox{\boldmath $[$}}
\def\br {\mbox{\boldmath $]$}}

\begin{document}

\title{Attractive Bose-Einstein condensates in anharmonic traps:
Accurate numerical treatment and the intriguing\\physics of the variance}
\author{Ofir E. Alon}
\email{ofir@research.haifa.ac.il}
\affiliation{Department of Mathematics, University of Haifa, Haifa 3498838, Israel}
\affiliation{Haifa Research Center for Theoretical Physics and Astrophysics, University of Haifa,
Haifa 3498838, Israel}
\author{Lorenz S. Cederbaum}
\email{lorenz.cederbaum@pci.uni-heidelberg}
\affiliation{Theoretische Chemie, Physikalisch--Chemisches Institut, Universit\"at Heidelberg, 
Im Neuenheimer Feld 229, D-69120 Heidelberg, Germany}

\begin{abstract}
The dynamics of attractive bosons trapped in one dimensional
anharmonic potentials is investigated.
Particular emphasis is put on 
the variance of the position and momentum many-particle operators.
Coupling of the center-of-mass and relative-motion degrees-of-freedom 
necessitates an accurate numerical treatment.
The multiconfigurational time-dependent Hartree for bosons (MCTDHB) method is used,
and high convergence of the energy, depletion and occupation numbers, and position and momentum variances 
is proven numerically.  
We demonstrate for the ground state and out-of-equilibrium dynamics,
for condensed and fragmented condensates,
for small systems and {\it en route} to the infinite-particle limit,
that intriguing differences between the density and variance of an attractive Bose-Einstein condensate 
emerge.
Implications are briefly discussed.
\end{abstract}


\maketitle 

\section{Introduction}\label{Intro}

Attractive bosons and Bose-Einstein condensates (BECs) 
have drawn considerable attention since long ago \cite{Mcguire_1964,Calogero_1975,frg1,ex2,Do_Attractive_1998,Muga_1998,Ueda_Leggett_1999,
Carr_2000,Castin_2001,sol1,sol2,
BMF,Ueda_2003,CCI0_2004,Ho_2004,Ueda_2004,Ring_2005,Buljan_2005,
Peter_DW_2006,Castin_2008,Frag_2008,
Fragility_2008,Peter_NJP_2008,Weiss_2009,Caton_2009,BJJ_Att_2010,Marios_2010,Polls_DW_2010,
Swift_2011,Castin_2012,Calabrese_2016,Drummond_2017,Malomed_2017,Reimann_2018}.
Attractive particles
tend to be closer together than non-interacting particles,
meaning that the shape of their density is narrower than the
respective density of non-interacting particles.
Being identical bosons
suggests that all particles could occupy one and the same single-particle state,
and be adequately described by Gross-Pitaevskii
mean-field theory.
Setting intuition aside,
it is needed to employ a many-body level of theory
in order to ascertain condensation and more so fragmentation of attractive bosons \cite{frg3}.
The latter has been found both in the case of statics, 
where the ground state can be fragmented due to symmetry \cite{Ueda_2003,CCI0_2004,Ho_2004,Ueda_2004},
and dynamics, where fragmentation can develop in time due to 
involvement of multi-particle excitations \cite{Frag_2008,Weiss_2009,Caton_2009,BJJ_Att_2010,Swift_2011},
see also \cite{Drummond_2017,Malomed_2017,Brand_Cos}.

For repulsive BECs the connection between condensation and the limit of an infinite-number of particles has drawn increased attention \cite{Castin_U,Yngvason_PRA,Lieb_PRL,Erdos_PRL,MATH_ERDOS_REF,
Variance,TD_Variance,SL_Psi,TD_Variance_BEC,L_new,Kaspar_Joerg,Var_Ani}.
There are rigorous results which
prove (when the product of the number of particles times the scattering length, i.e., the interaction parameter, is constant)
that the energy per particle and density per particle of the many-boson system coincide in the infinite-particle limit
with the respective Gross-Piteavekii quantities, 
and that the bosonic system is $100\%$ condensed
\cite{Yngvason_PRA,Lieb_PRL,Erdos_PRL,MATH_ERDOS_REF}, 
see also \cite{Castin_U,L_new}.
On the other hand, 
the variance of a many-particle operator,
such that the position, momentum, and angular-momentum operators \cite{Variance,TD_Variance,Kaspar_Joerg},
and the whole many-particle wavefunction \cite{L_new,SL_Psi}
can considerably deviate from those given by the Gross-Pitaevskii theory,
even in the infinite-particle limit when the 
bosonic system is $100\%$ condensed.
In particular, the position and momentum variances  
can behave in an opposite manner 
to the position and momentum densities
for repulsive bosons \cite{Variance,TD_Variance_BEC,Var_Ani}.
Indeed, the physics of the variance of many-particle
operators \cite{Drumm,Oriol_Robin} is much more involved than
that in the textbook case of a single particle \cite{QM_book}.

Less is known on the infinite-particle limit with attractive bosons,
except for a couple of solvable models \cite{Calogero_1975,hm6,hm3,EPJD,BB_HIM,hm15,BB_new}.
In comparison to their repulsive sibling, 
mathematical rigorous results
are more scarce \cite{proof_at1,proof_at2}.
Furthermore, and irrespective of the topic of the infinite-particle limit for attractive bosons, 
the variance of trapped attractive BECs has hardly been studied.
In the present work we investigate trapped attractive bosons in one-dimensional traps,
and focus on their many-particle position and momentum variances.
The above introductory remarks point toward the purpose of our work which is fourfold: 
(i) To discuss how the shape of an anharmonic trap in combination with inter-particle attraction impact the many-particle position and momentum variances;
(ii) To investigate the infinite-particle limit of a trapped attractive BEC, its degree of condensation, 
and whether and how the many-particle variances computed at the many-body level differ from those computed at the mean-field level, both for the ground state and for an out-of-equilibrium quench scenario;
(iii) To examine the many-particle variances of a fragmented attractive BEC
and in its infinite-particle limit;
(iv) To prove numerically high convergence of the energy, density, depletion, occupation numbers, and position and momentum variances 
for a variety of scenarios, from condensation to fragmentation, of trapped attractive BECs.

The structure of the paper is as follows.
In Sec.~\ref{Theory} we briefly 
discuss the variance in a many-body system 
and its computation from the wavefunction of a trapped BEC.
Our method of choice for the many-body wavefunction,
the multiconfigurational time-dependent Hartree for bosons (MCTDHB) method \cite{MCTDHB1,MCTDHB2}, 
is briefly discussed as well.
In Sec.~\ref{APPL} we present four detailed investigations,
two for condensed systems (Subsec.~\ref{sec_cond})
and two for fragmented (Subsec.~\ref{sec_frag}). 
Concluding remarks are put forward in Sec.~\ref{Conclusions}.
Finally,
further computational details and the discussion of convergence 
are collected in the Appendix.

\section{Theory}\label{Theory}

The many-body Hamiltonian of $N$ interacting bosons in one spatial dimension reads:
\beq\label{HAM}
 \hat H(x_1,\ldots,x_N;\lambda_0) = 
\sum_{j=1}^N \hat h(x_j) + \sum_{j<k} \lambda_0\hat W(x_j-x_k).
\eeq
Here, $\hat h(x) = -\frac{1}{2} \frac{\partial^2}{\partial x^2} + \hat V(x)$ is the one-particle Hamiltonian 
where $V(x)$ the trap potential and $W(x_1-x_2)$ the inter-particle interaction 
of strength $\lambda_0$.
Throughout this work the interaction in attractive,
i.e., $W(x)>0$ for $x<\infty$ and $\lambda_0<0$.
The units $\hbar=m=1$ are used. 
We investigate the ground state of the bosons,
\beq\label{TISE}
\hat H(x_1,\ldots,x_N;\lambda_0) \Phi(x_1,\ldots,x_N) =\break\hfill E \Phi(x_1,\ldots,x_N),
\eeq
for traps of different shapes as a function of the attraction strength $\lambda_0$.
$E$ is the total energy and $\Phi(x_1,\ldots,x_N)$ is normalized to one for all $N$.
For a quench of the attraction from $\lambda_0$ to $\lambda'_0<\lambda_0$, 
we solve the time-dependent Schr\"odinger equation, 
\beq\label{TDSE}
\hat H(x_1,\ldots,x_N;\lambda'_0) \Psi(x_1,\ldots,x_N;t) =
i \frac{\partial\Psi(x_1,\ldots,x_N;t)}{\partial t},
\eeq
with the initial condition being the ground state for $\lambda_0$, 
$\Psi(x_1,\ldots,x_N;0) = \Phi(x_1,\ldots,x_N)$.

Beyond energy,
other properties are needed to describe the system and interpret its properties.
Having the many-body wavefunction $\Psi(x_1,\ldots,x_N;t)$ at hand
allows one to
compute in principle any quantity of interest.
For our needs,
the reduced one-body density matrix
\beqn\label{1RDM}
\frac{\rho^{(1)}(x_1,x_1';t)}{N} &=&
\int dx_2 \ldots dx_N \, \Psi^\ast(x_1',x_2,\ldots,x_N;t) \Psi(x_1,x_2,\ldots,x_N;t) = \nonumber \\
 &=& \sum_j \frac{n_j(t)}{N} \, \alpha_j(x_1;t) \alpha^\ast_j(x'_1;t)
\eeqn
and the diagonal of the reduced two-body density matrix
\beqn\label{2RDM}
\frac{\rho^{(2)}(x_1,x_2,x_1,x_2;t)}{N(N-1)} &=& 
 \int dx_3 \ldots dx_N \, \Psi^\ast(x_1,x_2,\ldots,x_N;t) \Psi(x_1,x_2,\ldots,x_N;t) = \nonumber \\
 &=& \sum_{jpkq} \frac{\rho_{jpkq}(t)}{N(N-1)} \, 
\alpha^\ast_j(x_1;t) \alpha^\ast_p(x_2;t) \alpha_k(x_1;t) \alpha_q(x_2;t) 
\eeqn
are computed \cite{Lowdin,Yukalov,Mazz,RDMs}.
From the reduced one-particle density matrix the natural orbitals $\alpha_j(x;t)$, natural occupation numbers $n_j(t)$,
and the density of the system, $\rho(x;t) = \rho^{(1)}(x,x;t)$, are obtained.
In the reduced two-particle density matrix
the elements $\rho_{jpkq}(t) = \langle\Psi(t)|\hat b_j^\dag \hat b_p^\dag \hat b_k \hat b_q|\Psi(t)\rangle$
appear, where the creation $\hat b^\dag_j$ and annihilation $\hat b_j$ 
operators are associated with the single-particle functions $\alpha_j(x;t)$. 
It is convenient to enumerate the occupation numbers in order of non-increasing values.
Furthermore,
it is useful to call $\sum_{j>1}n_j(t)=N-n_1(t)$ the number of depleted particles, i.e., the depletion,
and $\frac{\sum_{j>1} n_j(t)}{N}=1-\frac{n_1(t)}{N}$ the depleted fraction. 
The latter are used to define the degree of condensation \cite{Penrose_Onsager}
or fragmentation \cite{frg1,frg3,frg2} of the bosonic system. 

Given the many-particle position operator, $\hat X=\sum_{j=1}^N \hat x_j$,
the variance per particle can be expressed as follows \cite{Variance,TD_Variance}:
\beqn\label{dis}
& & \frac{1}{N}\Delta_{\hat X}^2(t) = \frac{1}{N} 
\left[\langle\Psi(t)|\hat X^2|\Psi(t)\rangle - \langle\Psi(t)|\hat X|\Psi(t)\rangle^2\right] \equiv 
\Delta_{\hat x, density}^2(t) + \Delta_{\hat x, MB}^2(t), \nonumber \\
& & \quad \Delta_{\hat x, density}^2(t) = 
\int dx \frac{\rho(x;t)}{N} x^2 - \left[\int dx \frac{\rho(x;t)}{N} x \right]^2, \nonumber \\ 
& & \quad \Delta_{\hat x, MB}^2(t) = \frac{\rho_{1111}(t)}{N} \left[\int dx |\alpha_1(x;t)|^2 x \right]^2 
- (N-1) \left[\int dx \frac{\rho(x;t)}{N} x \right]^2 + \nonumber \\
& & \quad \quad + \sum_{jpkq\ne 1111} \frac{\rho_{jpkq}(t)}{N} \left[\int dx \alpha^\ast_j(x;t) \alpha_k(x;t) x \right]
\left[\int dx \alpha^\ast_p(x;t) \alpha_q(x;t) x\right]. \
\eeqn
The density term, $\Delta_{\hat x, density}^2(t)$,
describes the variance of $\hat x$, the single-particle position operator,
resulting from the shape of the density per particle $\frac{\rho(x;t)}{N}$.
The many-body term, $\Delta_{\hat x, MB}^2(t)$,
collects all other contributions to the many-particle variance of $\hat X$
emanating from correlations in the many-boson system.
Indeed,
$\Delta_{\hat x, MB}^2(t)$ identically equals to
zero within Gross-Pitaevskii theory.
Analogously, the variance per particle 
of the many-particle momentum operator
$\hat P_X=\sum_{j=1}^N \hat p_{x,j}$ is defined,
$\frac{1}{N}\Delta_{\hat P_X}^2(t)=\Delta_{\hat p_x, density}^2(t)+\Delta_{\hat p_x, MB}^2(t)$.

The center-of-mass position and momentum operators are defined as
$\hat X_{c.m.} = \frac{\hat X}{N}$ and $\hat P_{X_{c.m.}}=\hat P_X$ and satisfy
the usual commutation relation $\bl\hat X_{c.m.},\hat P_{X_{c.m.}}\br=i$ for any $N$.
Accordingly, the respective variances of the center-of-mass operators are given by
$\Delta^2_{\hat X_{c.m.}} = \frac{1}{N^2}\Delta^2_{\hat X}$
and $\Delta^2_{\hat P_{X_{c.m.}}} = \Delta^2_{\hat P_X}$
and satisfy the same uncertainty relation as the variances per particle of the
many-particle position and momentum operators do, 
$\Delta^2_{\hat X_{c.m.}}\Delta^2_{\hat P_{X_{c.m.}}} = 
\frac{1}{N}\Delta^2_{\hat X}\frac{1}{N}\Delta^2_{\hat P_X}$ \cite{Variance}.
This allows one to analyze the variances of bosonic systems in terms of both sets of many-particle operators
equivalently.

To investigate the ground state and out-of-equilibrium dynamics we 
recruit a suitable many-body theoretical and computational approach.
Out method of choice is the MCTDHB method \cite{MCTDHB1,MCTDHB2}
which has been extensively applied and documented in the literature 
\cite{BJJ,book_MCTDH,MCTDHB_OCT,Kaspar_The,MCTDHB_Shapiro,book_nick,LC_NJP,MCTDHB_3D_stat,
Peter_2015a,Axel_The,Peter_2015b,Uwe,Sven_Tom,Kota_2015,Peter_b,Alexej_u,
Kaspar_n,Axel_ar,Joachim,Axel_2018,Sudip,Alexej_2018}.
Briefly, the MCTDHB method represents the many-boson wavefunction as a linear combination
of all $\begin{pmatrix}N+M-1 \cr M-1\end{pmatrix}$
permanents $|\vec n;t\rangle$ generated by distributing the $N$ bosons over $M$ time-adaptive orbitals,
\beq
|\Psi(t)\rangle = \sum_{\vec n} C_{\vec n}(t) |\vec n;t\rangle,
\eeq
where $C_{\vec n}(t)$ are time-adaptive expansion coefficients and the vector $\vec n$ runs over
all the above $\begin{pmatrix}N+M-1 \cr M-1\end{pmatrix}$ distributions.
The time-adaptive orbitals and expansion coefficients are determined by 
equations-of-motion derived from the Dirac-Frenkel variational principle.
The MCTDHB equations-of-motion are propagated in imaginary \cite{MCHB} and real time
to compute the ground state (\ref{TISE}) and out-of-equilibrium dynamics (\ref{TDSE}), respectively. 
We use the numerical implementation in the software packages \cite{MCTDHB_LAB,package}.
There exist extensions of MCTDHB to more complex bosonic systems,
namely spinors \cite{MCTDHB_spin} and mixtures \cite{MCTDH_BB,ML1,ML2,ML3}.

The MCTDHB has been benchmarked in the literature \cite{Axel_The,Benchmarks} 
(also see \cite{Axel_MCTDHF_HIM,Cami_JPB})
with the exactly-solvable harmonic-interaction model \cite{hm6,hm3}. 
The interaction between the bosons in this benchmark is attractive and of infinite (long) range
(harmonic interaction).
In \cite{Kaspar_n}, the MCTDHB method was used
to reproduce the full counting distribution of the center-of-mass position operator,
$\hat X_{c.m.}$, of bosons in a harmonic trap
which is exactly solvable due to the separability of the center-of-mass
and relative-motion degrees-of-freedom.
The interaction between the bosons in this benchmark is also attractive but of zero range 
($\delta$-function interaction).
In \cite{BJJ_Att_2010}, the out-of-equilibrium dynamics of attractive (and repulsive, see also \cite{BJJ}) bosons,
interacting by a $\delta$-function interaction,
in a one-dimensional bosonic Josephson junction has been converged numerically.
In the investigations below, we model the interaction between
the bosons by an attractive Gaussian potential,
whose range is finite and obviously in between the ranges of attractions of the 
above benchmarks.
The Gaussian inter-particle interaction has often 
been used in the literature of ultra-cold bosons \cite{Peter_NJP_2008,Uwe,Kaspar_n,Axel_ar,Pathway_Pe,Peter_CoM_Rel_2B,SR_2009}.

\section{Results}\label{APPL}

We consider structureless bosons with a Gaussian attraction in a one spatial dimension.
The bosons are trapped in an anharmonic potential,
single well or a double well in the present study.
Because the trap is anharmonic,
the center-of-mass and relative-motion degrees-of-freedom 
are coupled \cite{Peter_CoM_Rel_2B}.
Consequently and unlike for harmonic traps, 
not even the many-particle position and momentum variances can be computed analytically,
thus necessitating a numerical treatment.
Furthermore, changing the interaction in the same trap is expected to alter the variances,
and examining different traps for the same interaction strength is expected
to alter the variances in a non-trivial manner.
All in all, the coupling between
the center-of-mass and relative-motion degrees-of-freedom
in the anharmonic trap {\it a-priori} implies that the variances depend on the shape of the trap and strength
of the attraction between particles.
This dependence is what we aim at studying in the first place.

\subsection{Condensation}\label{sec_cond}

\subsubsection{Ground state}\label{example1}

\begin{figure}[!]
\begin{center}
\vglue -0.5 truecm
\hglue -1.0 truecm
\includegraphics[width=0.345\columnwidth,angle=-90]{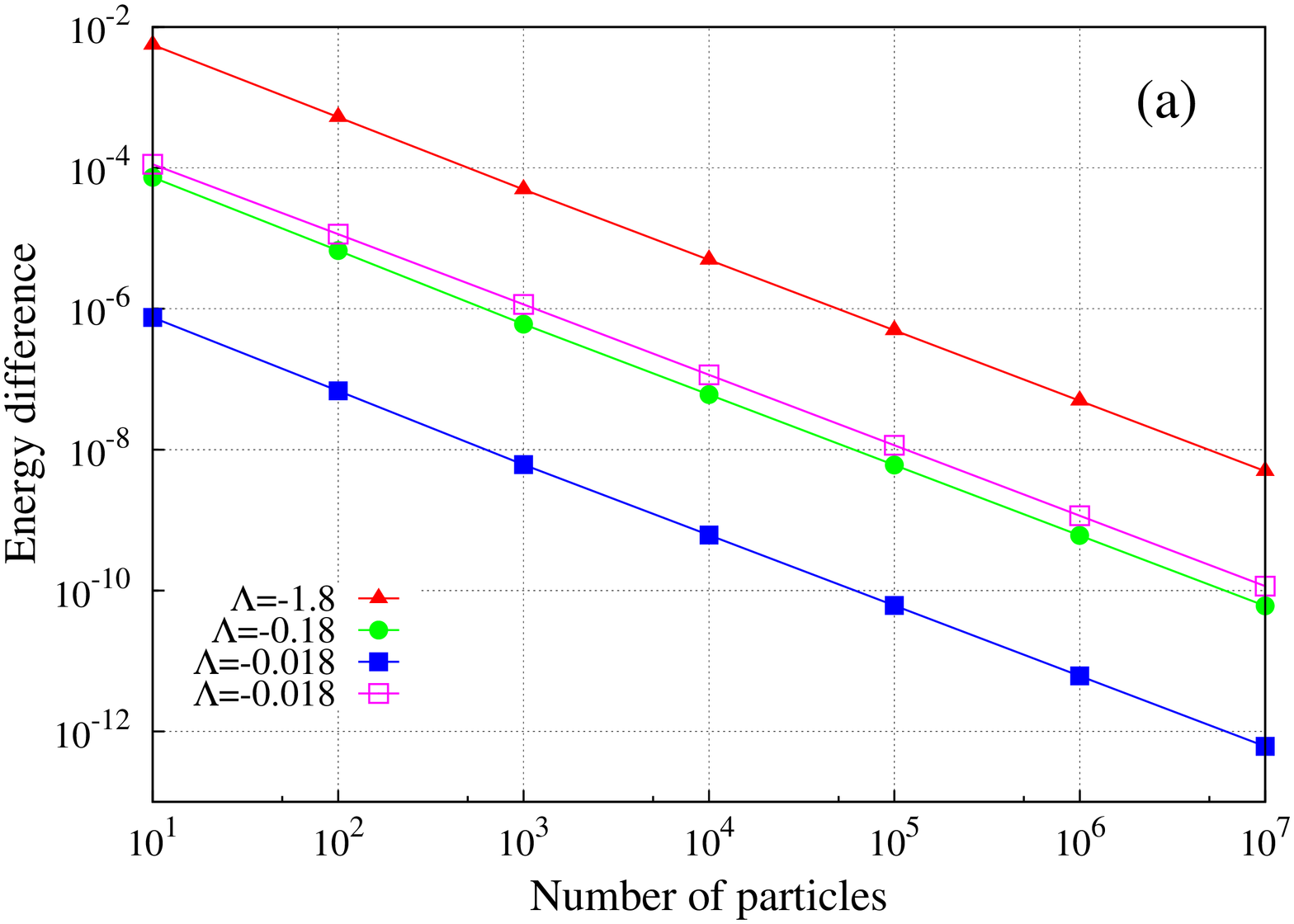}
\includegraphics[width=0.345\columnwidth,angle=-90]{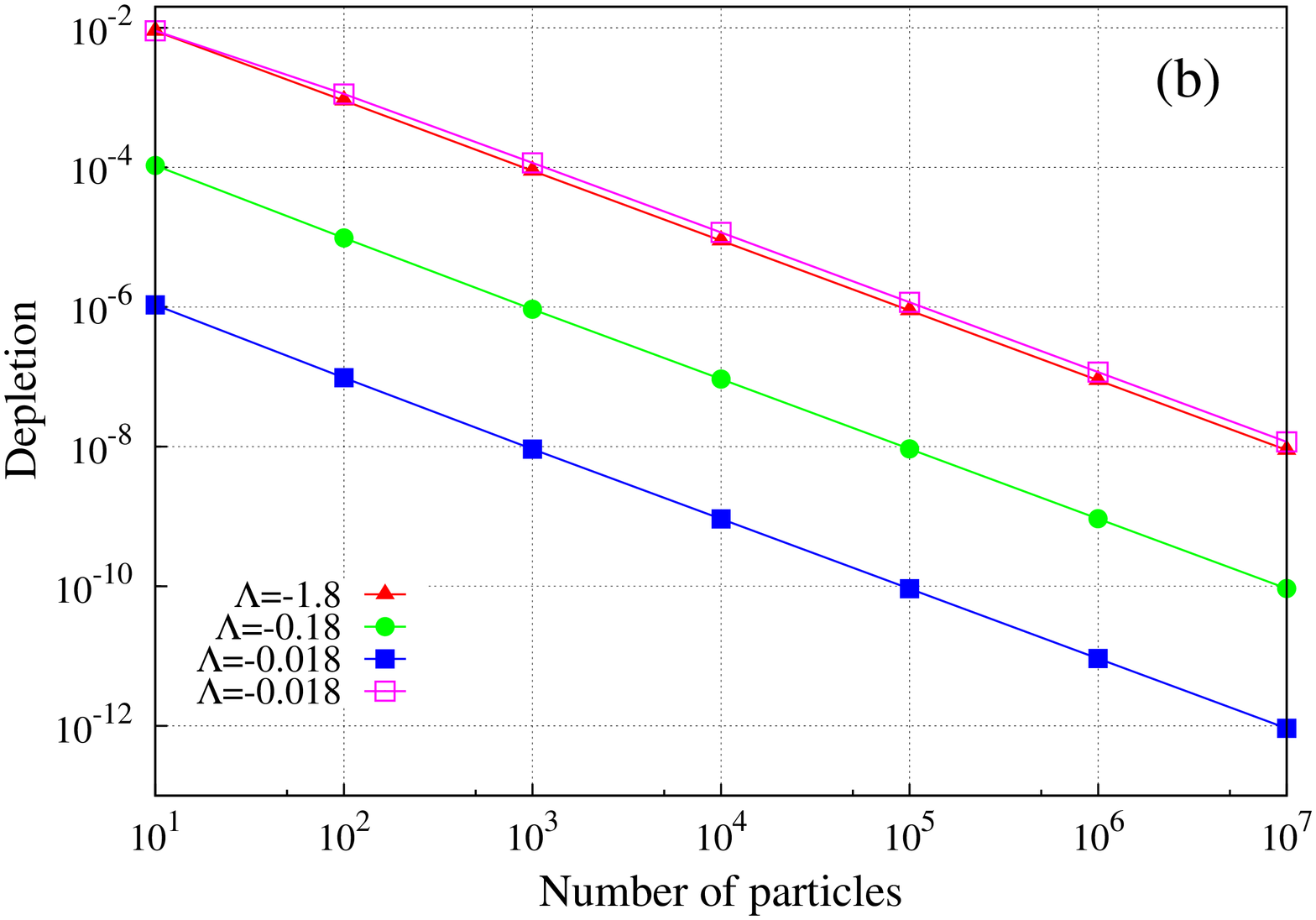}
\vglue 0.25 truecm
\hglue -1.0 truecm
\includegraphics[width=0.345\columnwidth,angle=-90]{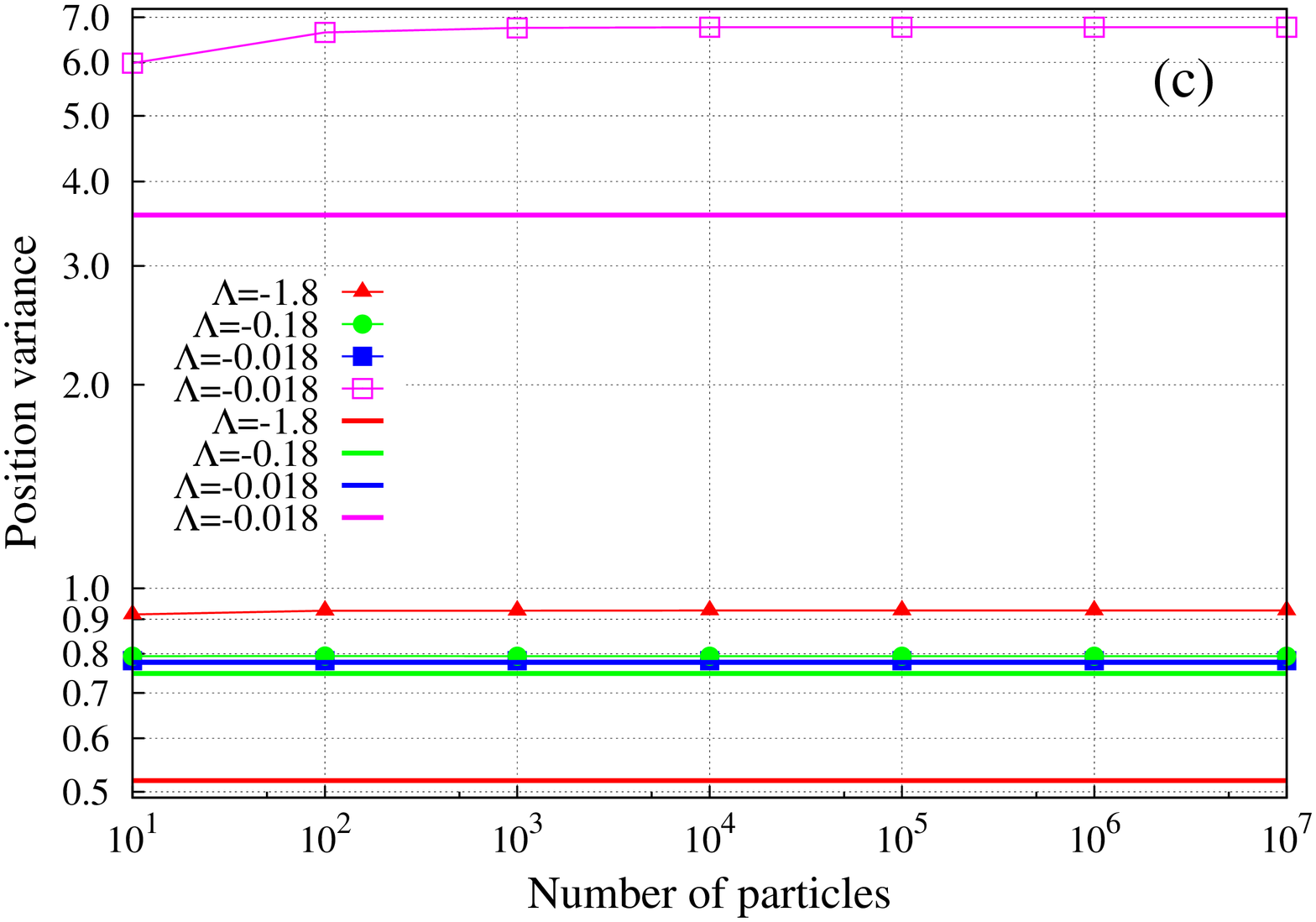}
\includegraphics[width=0.345\columnwidth,angle=-90]{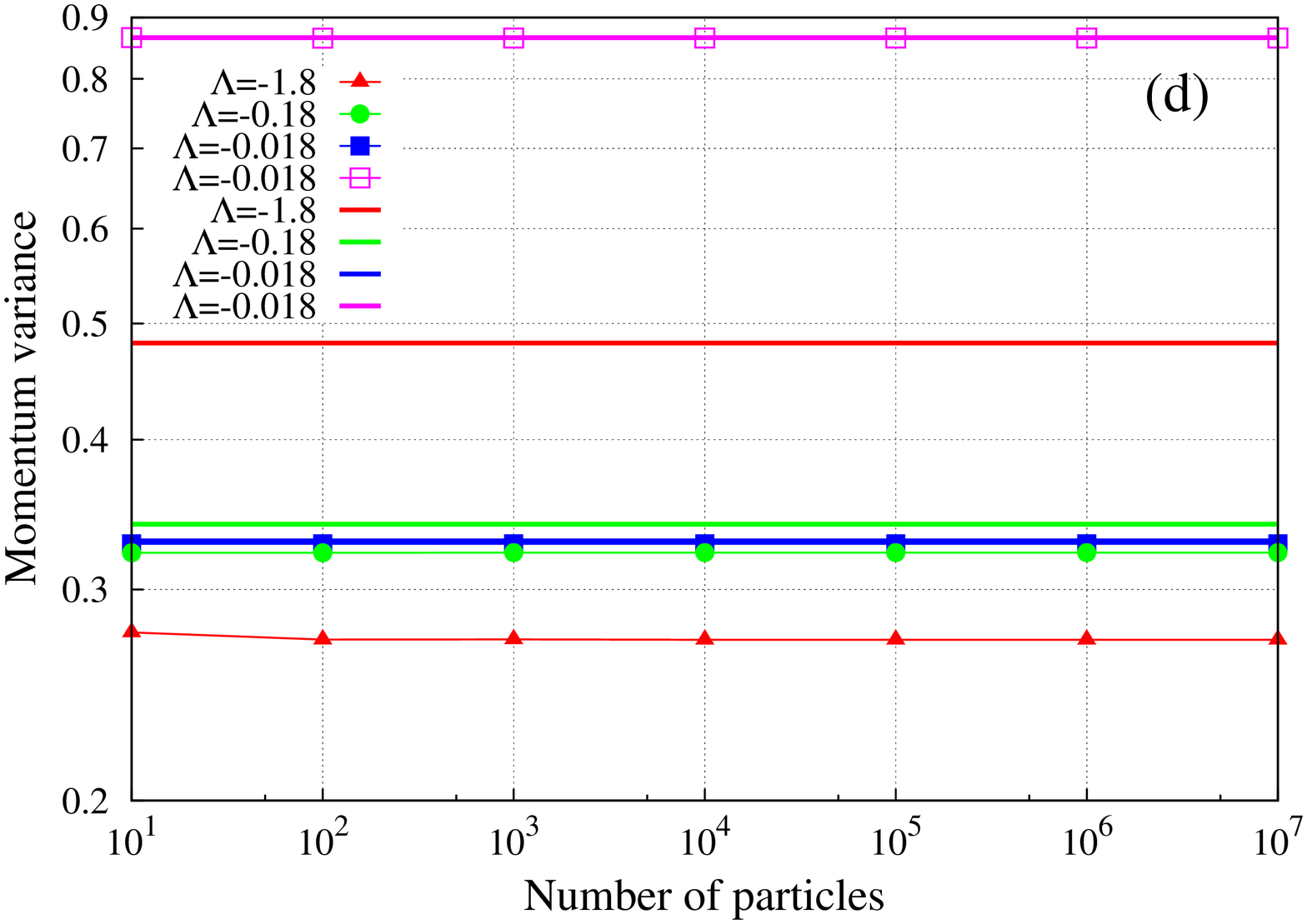}
\end{center}
\vglue 0.25 truecm
\caption{(Color online) 
Ground-state properties of an attractive trapped BEC: Condensation.
Shown as a function of the number of bosons $N$ are:
(a) Difference between the mean-field and many-body energies per particle;
(b) Depleted fraction, $1-\frac{n_1}{N}$;
(c) Many-particle position variance per particle, $\frac{1}{N}\Delta^2_{\hat X}$;
(d) Many-particle momentum variance per particle, $\frac{1}{N}\Delta^2_{\hat P_X}$.
The interaction parameters are $\Lambda=\lambda_0(N-1)=-0.018$, $\Lambda=-0.18$, and $\Lambda=-1.8$.
The confining potentials are the single well $V(x)=0.05x^4$ (filled symbols in red, green, and blue)
and the shallow double well $V(x)=0.05x^4+4.5e^{-0.5x^2}$ (open squares in magenta).
Many-body results in all panels are computed at the 
$M=5$ ($N=10$), $M=4$ ($N=100$), and
$M=2$ ($N\ge1000$) levels,
and marked by symbols
(the connecting lines are to guide the eye only); for demonstration of
accuracy and convergence see Fig.~\ref{f5}.
Mean-field variances (equivalent to $M=1$) are plotted in
panels (c) and (d) by smooth horizontal lines, 
being constant as a function of $N$.
The results suggest that the trapped attractive BECs are $100\%$ condensed
in the
infinite-particle limit, and that the position and momentum variances at the many-body level are, nonetheless,
different than at the mean-field level.
See the text for further discussion.
The quantities shown are dimensionless.}
\label{f1}
\end{figure}

We begin with the single-well anharmonic potential $V(x)=0.05x^4$.
The one-body Hamiltonian is $\hat h(x) = -\frac{1}{2}\frac{\partial^2}{\partial x^2} + 0.05x^4$.
The inter-particle interaction is the attractive Gaussian 
$\lambda_0 \hat W(x_1-x_2) = \lambda_0 e^{-0.5(x_1-x_2)^2}, \lambda_0<0$.
The range of the Gaussian is of no qualitative
consequence on the physics to be described below
(for a study with a narrower Gaussian see \cite{Peter_NJP_2008}),
but has the effect of accelerating the convergence of the computations with the number of orbitals $M$.
The interaction parameters are $\Lambda=\lambda_0(N-1)=-0.018$, $\Lambda=-0.18$, and $\Lambda=-1.8$.
The number of particles is $N=10, 100,\ldots,10\,000\,000$.
The number of orbitals is $M=5$ for $N=10$, $M=4$ for $N=100$, 
and $M=2$ for $N\ge1000$.
Fig.~\ref{f1} collects the results for the ground state.
The accuracy and convergence of the results are established in the Appendix,
see Fig.~\ref{f5}.

In Fig.~\ref{f1}a we plot and analyze the
difference between the mean-field and many-body energies per particle $\frac{E}{N}$.
The variational principle ensures this difference to be positive.
For all interaction strengths, the difference is diminished with increasing number of particles.
The results indicate that the many-body energy per particle approaches
from below the Gross-Pitaevskii energy per particle,
and provide strong numerical evidence that in the limit of an infinite number of particles the two energies coincide.
In Fig.~\ref{f1}b we depict the depleted fraction $1-\frac{n_1}{N}$.
As the number of particles increases,
the depleted fraction reduces towards $0\%$, i.e.,
the condensate fraction increases towards $100\%$.
Again, the results provide strong numerical evidence that,
in the infinite-particle limit,
the condensate fraction of a trapped attractive BEC 
(in one spatial dimension) is $100\%$.

In Fig.~\ref{f1}c we plot the many-particle position variance per particle, $\frac{1}{N}\Delta^2_{\hat X}$,
and in Fig.~\ref{f1}d the corresponding momentum variance, $\frac{1}{N}\Delta^2_{\hat P_X}$.
Here, as the number of bosons increases and at constant interaction parameters,
the variances computed at the many-body level saturate
at values different than those computed 
at the mean-field level.
To be specific, as the attraction increases the position (momentum) variance decreases (increases) at the mean-field level, which is in line with narrowing of the density.
Yet and counterintuitively, the many-body quantities exhibit
the inverse behavior, namely, the position (momentum) variance increases (decreases) with
increasing attraction strength.
We find at the mean-field level $\frac{1}{N}\Delta^2_{\hat X}=\Delta_{\hat x, density}^2=0.777, 0.748, 0.519$ and 
$\frac{1}{N}\Delta^2_{\hat P_X}=\Delta_{\hat p_x, density}^2=0.329, 0.340, 0.482$,
whereas at the many-body level we find
$\frac{1}{N}\Delta^2_{\hat X}=0.781, 0.794, 0.927$ and
$\frac{1}{N}\Delta^2_{\hat P_X}=0.328, 0.322, 0.272$
for the three interactions parameters 
$\Lambda=-0.018, -0.18, -1.8$, respectively.
Just like for their repulsive sibling \cite{Variance,TD_Variance_BEC,Var_Ani},
the variance and density in the ground state
of an attractive trapped BEC behave in an opposite manner.
In summary, increasing the attraction amounts to enlarging the position variance, despite narrowing of the density,
in as much as increasing the repulsion \cite{Variance,TD_Variance_BEC,Var_Ani} leads to decreasing of the position variance, in spite of 
broadening of the density. 

For comparison with the single-well potential $V(x)=0.05x^4$ discussed above, 
the shallow double-well potential $V(x)=0.05x^4+4.5e^{-0.5x^2}$ is examined
for the weakest interaction parameter $\Lambda=-0.018$, see Fig.~\ref{f1}.
The shallow double well can be seen as a further distortion of the anharmonic single-well potential.
Examining the energy difference, Fig.~\ref{f1}a,
and depletion, Fig.~\ref{f1}b,
again indicates  
coincidence of the many-body quantities in the infinite-particle limit with the Gross-Pitaevskii quantities,
namely $100\%$ condensation.
For the variances, the many-body $\frac{1}{N}\Delta^2_{\hat X}=6.762$
and mean-field $\frac{1}{N}\Delta^2_{\hat X}=\Delta_{\hat x, density}^2=3.566$
position variances show a larger difference than in the single-well trap for the same interaction parameter;
Compare the magenta and blue curves in Fig.~\ref{f1}c.
The reason is the larger depletion
of the BEC in the double well than in the single well 
for the same interaction parameter, see Fig.~\ref{f1}b.
On the other hand, the many-body $\frac{1}{N}\Delta^2_{\hat P_X}=0.865$
and mean-field $\frac{1}{N}\Delta^2_{\hat P_X}=\Delta_{\hat p_x, density}^2=0.866$
momentum variances are hardly distinguishable in the double well
(just like the corresponding quantities in the single well for the same interaction parameter $\Lambda=-0.018$),
despite the larger depletion. 
This is because the small overlap between the orbitals and their spatial derivative
(coming from the momentum operator $\hat p_x$) entering the many-body term 
$\Delta_{\hat p_x, MB}^2$ of the variance (\ref{dis}),
see in this context \cite{TD_Variance_BEC}.
 
\subsubsection{Out-of-equilibrium dynamics}\label{example2}

So far we have investigated the many-particle variance of the
ground state and seen how it depends on the strength of the attraction and shape of the trap.
Although the BECs studied above are 
essentially fully condensed, the position and momentum variances behave oppositely to the
respective densities at the many-body level of theory.
This is in contrast to the mean-field behavior.

\begin{figure}[!]
\begin{center}
\vglue -2.0 truecm
\hglue -1.0 truecm
\includegraphics[width=0.295\columnwidth,angle=-90]{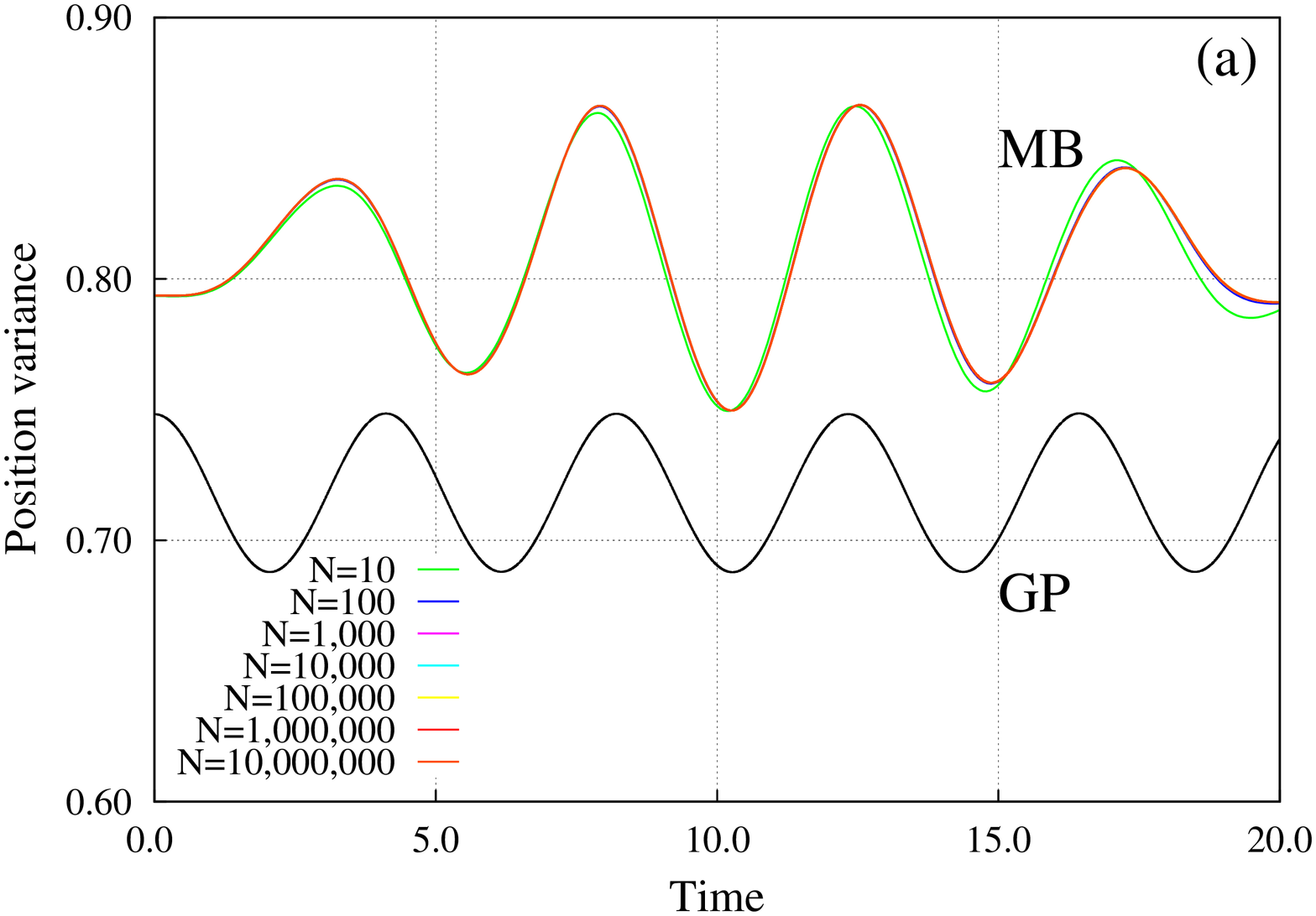}
\vglue 0.15 truecm
\hglue -1.0 truecm
\includegraphics[width=0.295\columnwidth,angle=-90]{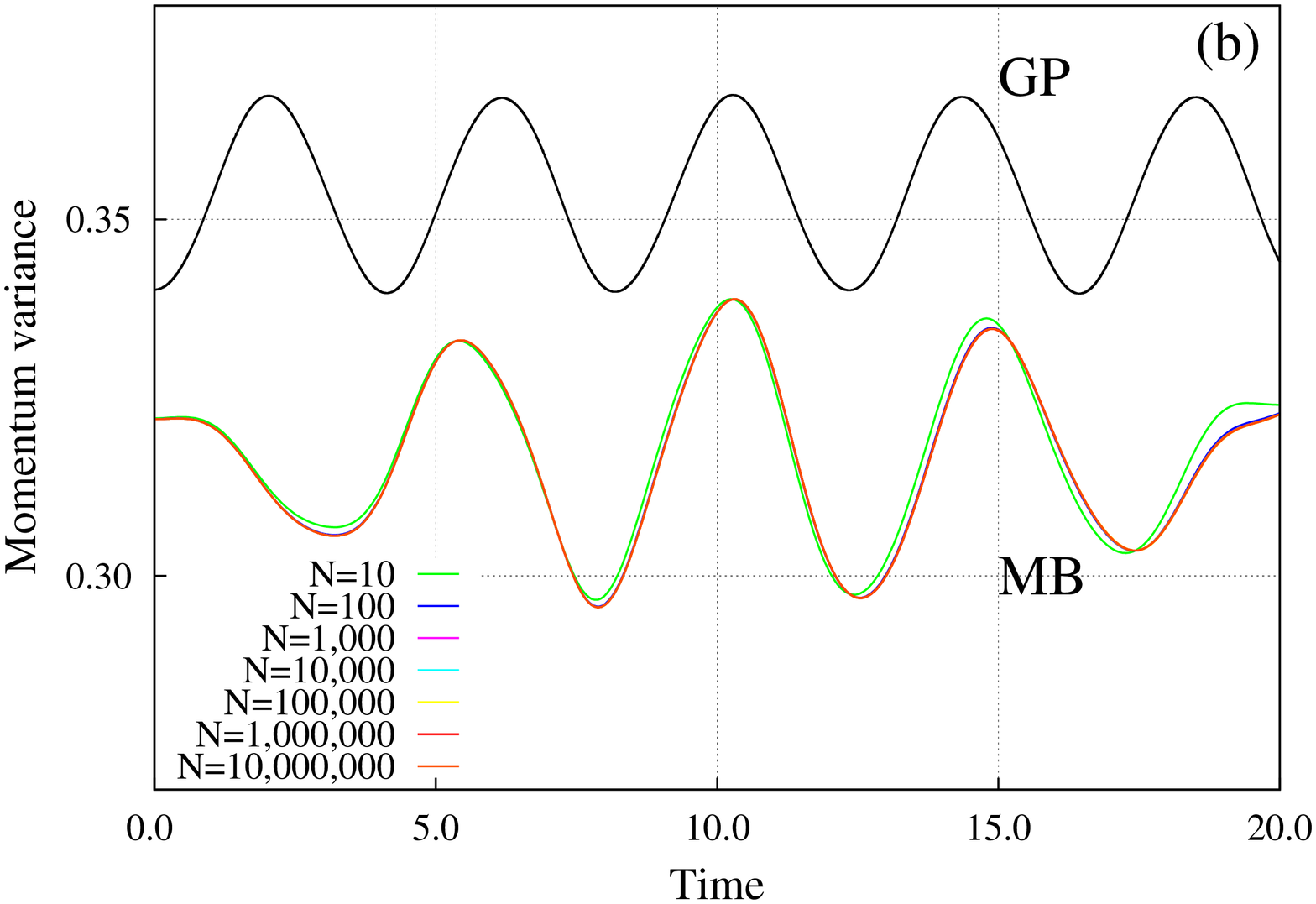}
\vglue 0.15 truecm
\hglue -1.0 truecm
\includegraphics[width=0.295\columnwidth,angle=-90]{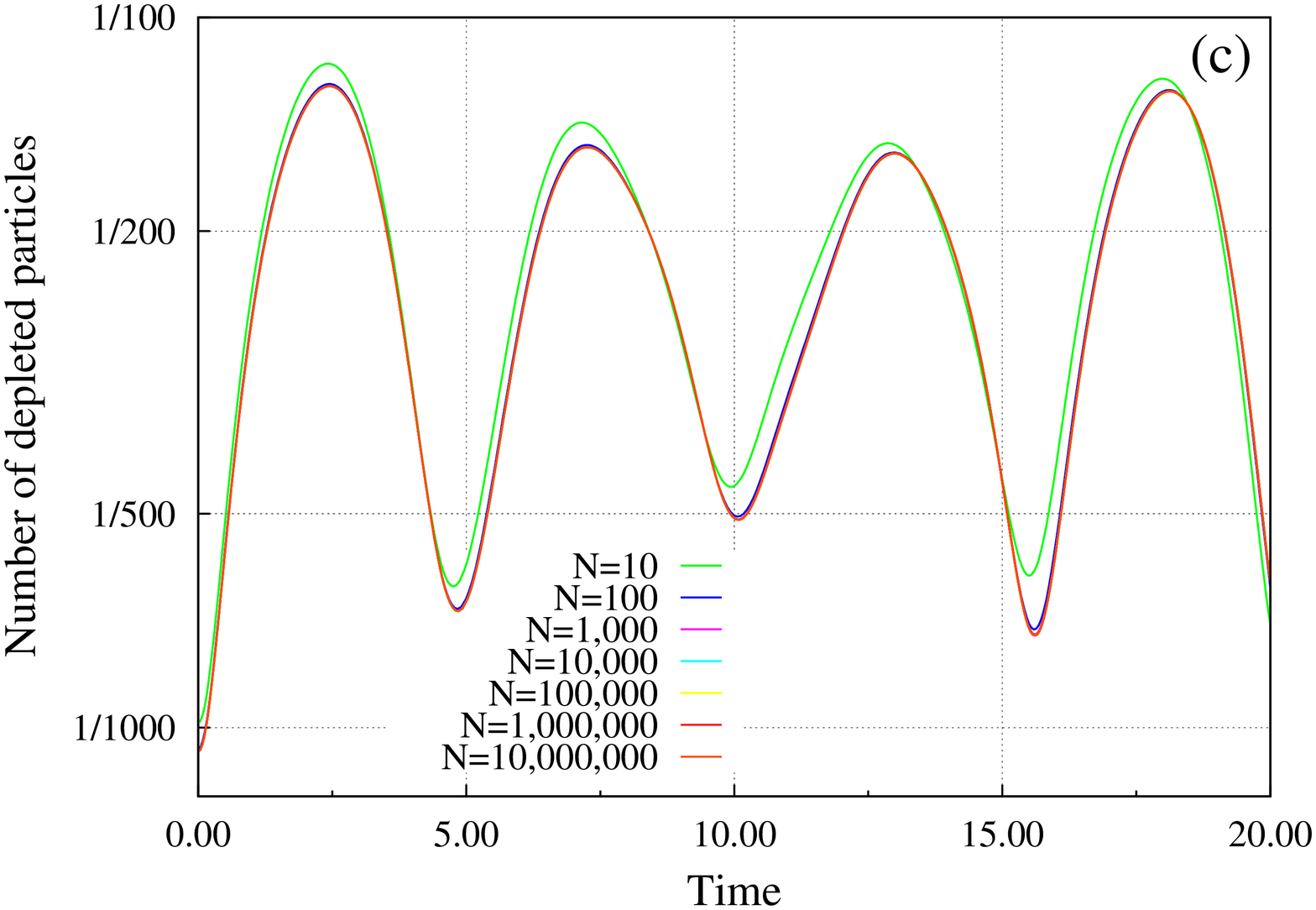}
\end{center}
\vglue 0.15 truecm
\caption{(Color online)
Breathing dynamics of an attractive trapped BEC following an interaction quench.
Shown as a function of the number of bosons $N$ are:
(a) Many-particle position variance per particle, $\frac{1}{N}\Delta^2_{\hat X}(t)$;
(b) Many-particle momentum variance per particle, $\frac{1}{N}\Delta^2_{\hat P_X}(t)$;
(c) Number of depleted particles, $N-n_1(t)$.
The interaction parameter is quenched from $\Lambda=\lambda_0(N-1)=-0.18$
to $\Lambda=-0.36$ at $t=0$.
The confining potential is the single well $V(x)=0.05x^4$.
Many-body results in all panels are computed at the $M=2$ level;
For demonstration of accuracy and convergence see Fig.~\ref{f6}.
Mean-field results (equivalent to $M=1$)
are plotted for comparison [in panels (a) and (b)]
by black lines for all $N$.
At the many-body level the position (momentum) variance initially increases (decreases),
oppositely to the respective mean-field variances.
The results suggest that the out-of-equilibrium 
trapped attractive BEC is $100\%$ condensed (the depleted fraction is $0\%$)
in the infinite-particle limit, 
and that the time-dependent position and momentum variances at the many-body level 
behave differently from those at the mean-field level.
Here, they can differ by more than $10\%$ despite having less than $\frac{1}{100}$-th of a particle depleted
for any $N$.
See the text for further discussion.
The quantities shown are dimensionless.}
\label{f2}
\end{figure}

It is instructive to examine an out-of-equilibrium scenario and inquire, similarly,
whether a trapped attractive BEC remains $100\%$ condensed 
in the infinite-particle limit and
whether the time-dependent variances behave counterintuitively as well.
For this, we consider
a quench scenario in the single-well anharmonic potential $V(x)=0.05x^4$.
The one-body Hamiltonian is $\hat h(x) = -\frac{1}{2}\frac{\partial^2}{\partial x^2} + 0.05x^4$.
The interaction is quenched at $t=0$
from $\Lambda=\lambda_0(N-1)=-0.18$ to $\Lambda=-0.36$.
Following the quench of the interaction, 
the density performs breathing oscillations \cite{TD_Variance_BEC,Var_Ani,Bonz,Peter_2013,MCTDHB_3D_dyn}.
The number of particles is $N=10, 100,\ldots,10\,000\,000$ and the number of orbitals is $M=2$.
Fig.~\ref{f2} collects the results of the out-of-equilibrium dynamics.
The accuracy and convergence of the results are established in the Appendix,
see Fig.~\ref{f6}.

In Fig.~\ref{f2}a we plot the many-particle position variances per particle, $\frac{1}{N}\Delta^2_{\hat X}(t)$,
at the many-body and mean-field levels,
and in Fig.~\ref{f2}b the respective momentum variances, $\frac{1}{N}\Delta^2_{\hat P_X}(t)$, are depicted.
The many-body curves quickly overlap each other as the number of bosons $N$ is increased
at the constant interaction parameters of the quench.
Furthermore,
they can differ by more than $10\%$ from the mean-field curve,
presenting the results for any $N$.
This sizable difference emerges due to a tiny number of depleted particles,
see discussion below.
All variances vary in time in an oscillatory manner,
signifying the breathing of the cloud following the interaction quench.
Importantly, the time-dependent many-body variances behave in an opposite manner to
the mean-field variances.
Indeed, making the attraction suddenly stronger,
the position (momentum) density initially narrows (broadens),
as do the Gross-Pitaevskii variances show.
Yet, the many-body position (momentum) variance initially actually increases (decreases),
i.e., behaves in an opposite manner to the density.

In Fig.~\ref{f2}c the number of depleted particles, $N-n_1(t)$, is shown as a function of time.
For any number $N$, even for $10\,000\,000$ bosons, there is less than a $\frac{1}{100}$-th of a boson
outside the condensed mode.
Furthermore, the time-dependent depletion quickly saturates with $N$, i.e.,
the respective curves overlap each other.
These constitute strong numerical evidence that the out-of-equilibrium attractive trapped
BEC becomes $100\%$ condensed (the depleted fraction becomes $0\%$) in the infinite-particle limit.

\subsection{Fragmentation}\label{sec_frag}

\subsubsection{Finite condensate}\label{example3}

\begin{figure}[!]
\begin{center}
\vglue -0.5 truecm
\hglue -1.0 truecm
\includegraphics[width=0.345\columnwidth,angle=-90]{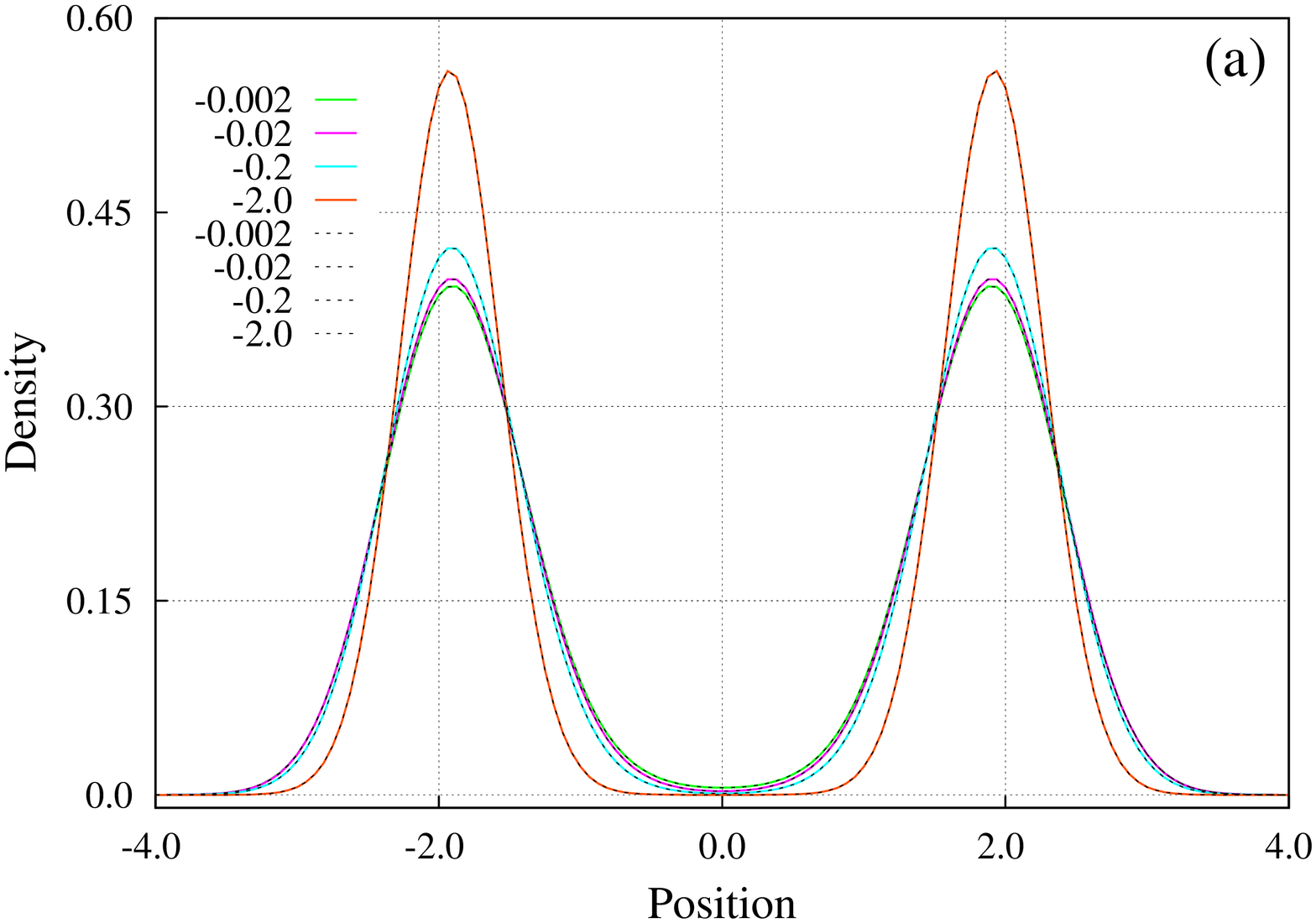}
\includegraphics[width=0.345\columnwidth,angle=-90]{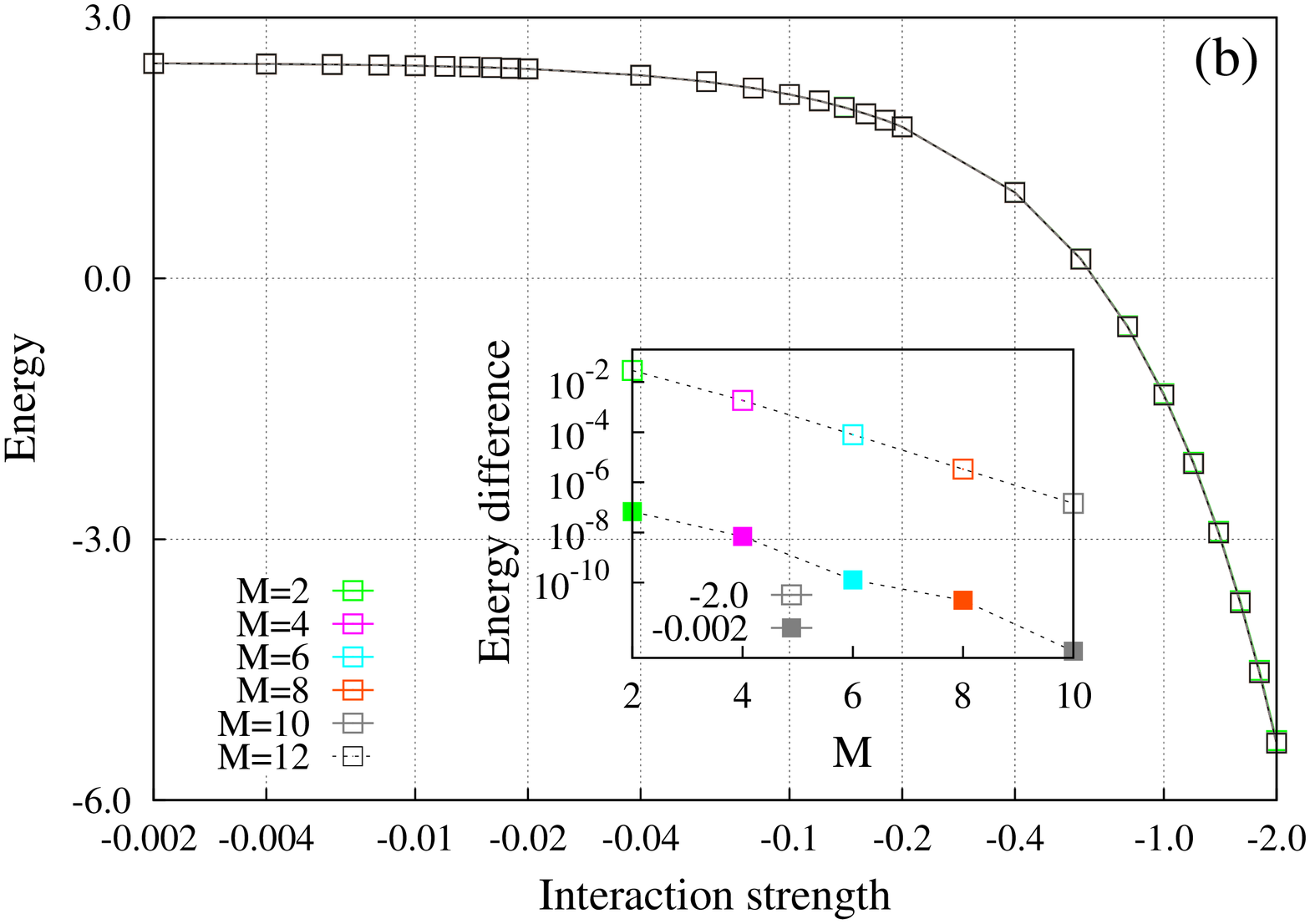}
\vglue 0.25 truecm
\hglue -1.0 truecm
\includegraphics[width=0.345\columnwidth,angle=-90]{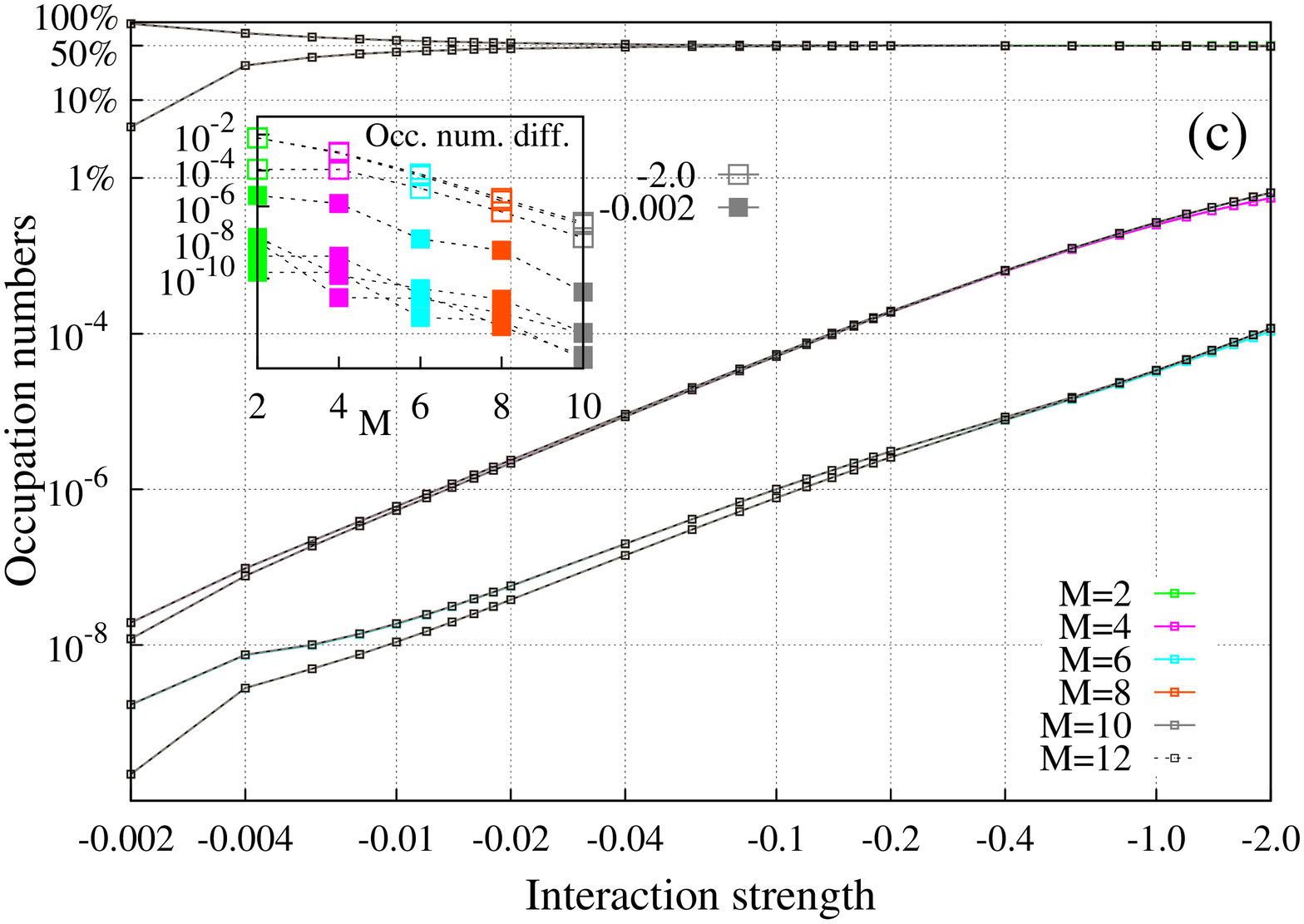}
\includegraphics[width=0.345\columnwidth,angle=-90]{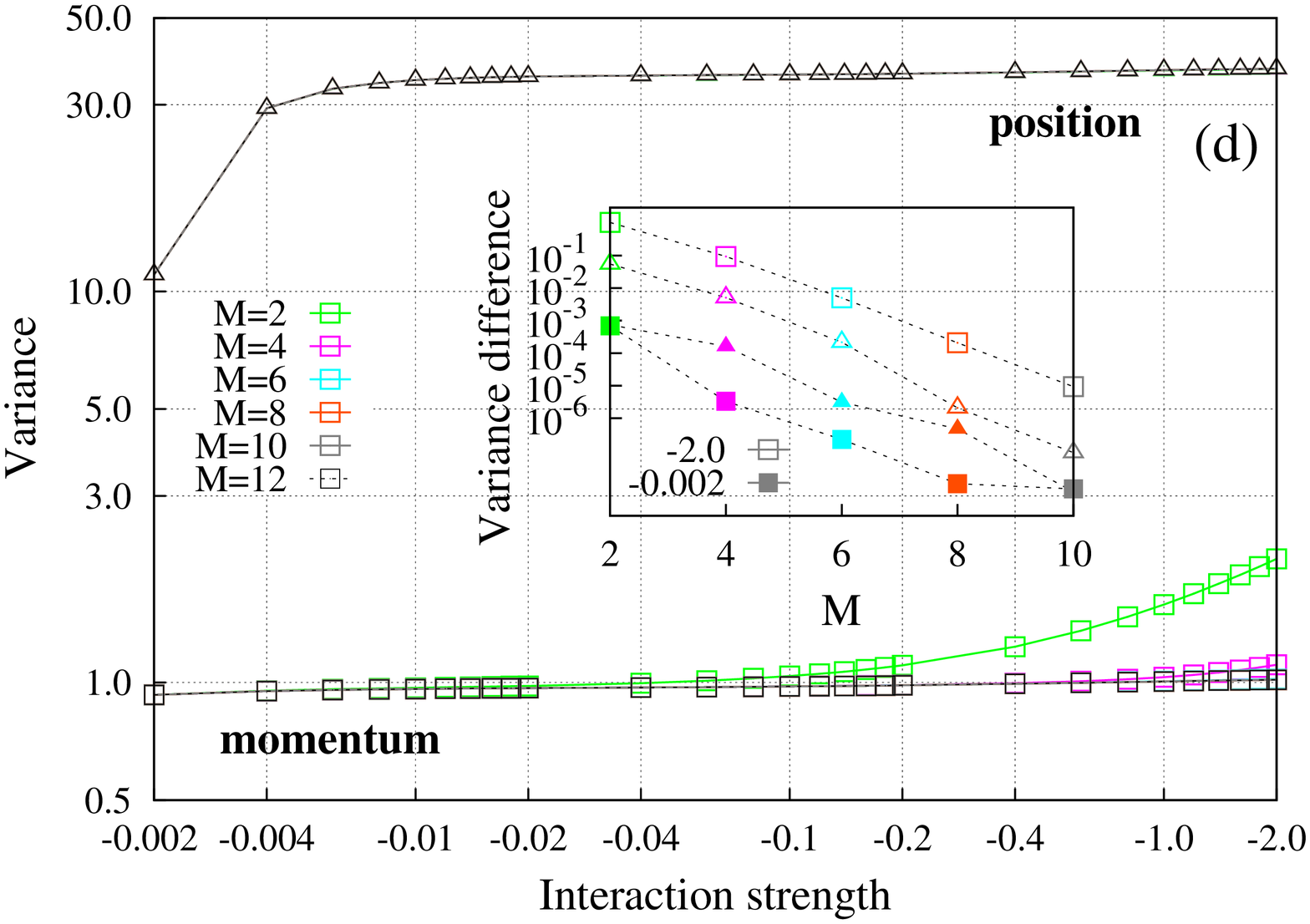}
\end{center}
\vglue 0.25 truecm
\caption{(Color online)
Ground-state properties of an attractive trapped BEC along the pathway from condensation to fragmentation.
Shown for $N=10$ bosons are:
(a) Density per particle, $\frac{\rho(x)}{N}$;
(b) Energy per particle, $\frac{E}{N}$;
(c) Occupation numbers per particle, $\frac{n_j}{N}, j=1,\dots,6$ in percents;
(d) Many-particle position variance per particle, $\frac{1}{N}\Delta^2_{\hat X}$,
and momentum variance per particle, $\frac{1}{N}\Delta^2_{\hat P_X}$.
The interaction strengths are $\lambda_0=-0.002,-0.004,\ldots,-1.8,-2.0$.
The confining potential is the double well $V(x)=0.05x^4+5.0e^{-0.5x^2}$.
In panel (a), the densities computed with $M=10$ (solid curves in colors) and
with $M=12$ (dashed curves in black) self-consistent orbitals are plotted
(for the four of interaction strengths $-0.002,-0.02,-0.2,-2.0$) and seen to lie atop each other.
The results depicted in panels (b)-(d) are computed at the $M=2,\dots,12$ level and marked by
symbols connected by lines to guide the eye.
The insets provide the absolute differences of the results computed at the various
$M<12$ levels to those at the final $M=12$ level for two interaction strengths ($-0.002$ and $-2.0$).
The color palettes of the insets and panels are the same.
The data shown proves the high numerical convergence of the results.
See the text for further discussion.
The quantities shown are dimensionless.}
\label{f3}
\end{figure}

Fragmentation of BECs has drawn a broad activity \cite{frg3,Sipe,ALN,AL_2005,Pathway,Pathway_Pe,Uwe_PRL1,frg4,frg5,frg6,frg7,Uwe_PRL2,frg8,Uwe_PRL3,frg9}.
Here, our first investigation is of a finite and small system of attractive bosons in the ground
state of a double-well potential.
In particular, we examine the many-body variances along the
pathway from condensation to fragmentation of the attractive bosons
in a double well \cite{Ho_2004}.
What should we expect?
Fig.~\ref{f3} collects the results.

The confining potential is the double well $V(x)=0.05x^4+5.0e^{-0.5x^2}$,
and the number of bosons is $N=10$.
The interaction strengths are $\lambda_0=-0.002,-0.004,\ldots,-1.8,-2.0$,
and the number of orbitals used in the computations is $M=2,4,\dots,12$.
In Fig.~\ref{f3}a we depict the ground-state density per particle, $\frac{\rho(x)}{N}$,
for four of the above interaction strengths ($-0.002$, $-0.02$, $-0.2$, and $-2.0$).
As can be seen, the density narrows with the increase of the attraction.
This signifies coupling to higher bands in the double well.
In Fig.~\ref{f3}b the energy per particle, $\frac{E}{N}$, is depicted.
The ground-state energy of the bosons is, of course, lowered by increasing the attraction.

Side by side with narrowing of the density and lowering of the energy,
the increase in the attraction leads to fragmentation of the system.
Fig.~\ref{f3}c records the six largest occupation
numbers per particle of the bosonic system,
$\frac{n_j}{N}, j=1,\ldots,6$.
For the weakest attraction, $\lambda_0=-0.002$, 
we find $\frac{n_1}{N}=95.46\%$, $\frac{n_2}{N}=4.54\%$, and
$\frac{n_{j\ge3}}{N}<3 \times 10^{-8}$. 
For the strongest interaction, $\lambda_0=-2.0$,
we have $\frac{n_1}{N}=\frac{n_2}{N}=49.34(3)\%$, $\frac{n_3}{N}=\frac{n_4}{N}=0.64(5)\%$, 
and $\frac{n_5}{N}=\frac{n_6}{N}=0.01(2)\%$.
The computation confirms in a continuous double-well system,
taking all contributing bands into account,
that the attractive bosons have become two-fold fragmented.

In Fig.~\ref{f3}d we prescribe the many-body position and momentum variances as 
a function of the attraction strength $\lambda_0$.
We first discuss the results qualitatively and thereafter quantitatively. 
The position variance is already larger than $10$ for the weakest attraction,
and it surpasses $35$ for the strongest attraction.
For comparison,
the system's physical size can be inferred from the shape of the density in Fig.~\ref{f3}a,
and is smaller than $5$.
Thus, the behavior we have seen for condensed attractive bosons in Fig.~\ref{f1},
that already the smallest depletion leads to a larger position variance than the
mean-field variance (equivalent to the shape of the density),
persists and more so is enhanced for fragmented attractive bosons.
With less than $5\%$ fragmentation the many-body variance is more than twice the system's physical size,
and with nearly $50\%$ fragmentation it is about eight-times the system's physical size.
At the other end, despite the density getting narrower and the bosons fragmented,
the many-particle momentum variance increases only very mildly and monotonically, 
see Fig.~\ref{f3}d.

The convergence of the system's properties collected in Fig.~\ref{f3} deserves a separate discussion.
There are two issues here, that different properties converge at a different pace with
increasing number of self-consistent orbitals $M$ and, of course, 
the particular level of convergence that can be achieved.

We begin with the densities per particle, $\frac{\rho(x)}{N}$, plotted in Fig.~\ref{f3}a.
For all four interactions strengths, $\lambda_0=-0.002,-0.02,-0.2,-2.0$,
i.e., from weak to strong attraction (from small to large fragmentation, see Fig.~\ref{f3}c),
$\frac{\rho(x)}{N}$ with $M=10$ self-consistent orbitals are indistinguishable 
from those with $M=12$.
The profiles of the respective densities precisely
lie atop each other.

Next, the inset in Fig.~\ref{f3}b follows the convergence of the energy per particle, $\frac{E}{N}$,
by plotting the differences of the results computed at the various
$M<12$ levels to those at the final $M=12$ level 
(the variational principle ensures these differences to be positive)
for two interaction strengths.
For the weakest interaction strength, $\lambda_0=-0.002$,
the energy per particle is found to converge
to better than $10^{-10}$,
whereas for the strongest interaction strength, $\lambda_0=-2.0$,
to better than $10^{-5}$.
Looking at the main panel, Fig.~\ref{f3}b,
this pace of convergence explains why already the results with $M\ge 6$ visually lie
atop each other
for the entire range of interaction strengths.   

Then, the inset in Fig.~\ref{f3}c follows the convergence of the occupation numbers per particle,
$\frac{n_j}{N}, j=1,\ldots,6$
by plotting the absolute differences of the results computed at the various
$M<12$ levels to those at the final $M=12$ level.
We find that all occupation numbers converge to better than $10^{-8}$ for the
weakest interaction strength, and better than $10^{-5}$ for the strongest interaction strength.
Looking at the main panel, Fig.~\ref{f3}c,
this pace of convergence explains why already the results with $M\ge 4$ orbitals
visually lie
atop each other for the two largest occupation numbers $\frac{n_1}{N}$ and $\frac{n_2}{N}$,
for the entire range of interaction strengths.
Similarly, the results with $M\ge 6$ visually lie
atop each other for $\frac{n_3}{N}$ and $\frac{n_4}{N}$,
and the results with $M\ge 8$ visually lie
atop each other for $\frac{n_5}{N}$ and $\frac{n_6}{N}$.   
It is useful to stress how the convergence of the individual occupation numbers is assessed \cite{Accurate}.
Convergence is determined from `top to bottom' when increasing the number of self-consistent orbitals $M$,
namely, from the largest occupation number per particle, $\frac{n_1}{N}$,
down to the smallest occupation number per particle, $\frac{n_6}{N}$.
  
Last but not least is the convergence of
the many-particle position variance per particle and momentum variance
per particle shown in the inset of Fig.~\ref{f3}d.
Again, absolute differences of the results computed at the various
$M<12$ levels to those at the final $M=12$ level are plotted.  
We find that the position variance per particle is converged to better than $10^{-6}$
for the weakest $\lambda_0=-0.002$ as well as the strongest $\lambda_0=-2.0$ interaction
strengths.
The momentum variance per particle, the most sensitive property of the BEC discussed here,
is converged for the weakest interaction strength to better than $10^{-6}$,
and for the strongest interaction
strength to better than $10^{-4}$.
Examining the main panel, Fig.~\ref{f3}d,
shows that the results with $M\ge 4$ orbitals
visually lie atop each other for $\frac{1}{N}\Delta^2_{\hat X}$,
and the results with $M\ge 8$ orbitals 
visually lie atop each other
for $\frac{1}{N}\Delta^2_{\hat P_X}$.
All in all,
for the many-body variances corresponding to the four densities in Fig.~\ref{f3}a
($\lambda_0=-0.002,-0.02,-0.2,-2.0$), 
we determine
$\frac{1}{N}\Delta^2_{\hat X}=11.0339(5), 35.4209(7), 36.0365(6), 37.1024(2)$
for the position variance per particle
and $\frac{1}{N}\Delta^2_{\hat P_X}=0.9292(4), 0.9671(9), 0.9830(1), 1.0156(2)$
for the momentum variance per particle, see Fig.~\ref{f3}d
(last digit of the momentum variance for the strongest attraction 
was verified by a computation using $M=14$ orbitals).
Indeed,
high convergence of the many-particle position and momentum variances of an attractive
trapped BEC is achieved and demonstrated for the entire pathway from condensation to fragmentation.

\subsubsection{Condensate {\it en route} to the infinite-particle limit}\label{example4}

Our final investigation draws upon the above results
and a fundamental difference between
the fragmented ground state of attractive and repulsive trapped bosons.
To remind, 
let the ground state of a finite repulsive BEC, 
consisting of $N$ particles in a trap and for a certain interaction parameter $\Lambda=\lambda_0(N-1)$,
be fragmented.
Then, increasing $N$ while keeping $\Lambda$ constant, 
the fragmentation eventually diminishes and 
the repulsive BEC becomes condensed for a rather finite $N$ \cite{AL_2005}.
The repulsive BEC becomes precisely $100\%$ condensed in the infinite-particle limit \cite{Lieb_PRL}. 
   
For a trapped attractive BEC the situation can differ,
since symmetry plays a decisive role.
Briefly, in a symmetric trap, say a double well or a ring,
the ground state becomes fragmented from a certain attraction strength on
at the many-body (or, beyond Gross-Pitaevskii) level of theory \cite{BMF,Ueda_2003,Ho_2004,CCI0_2004}.
There are fingerprints for this fragmentation at the mean-field level of theory,
when two solutions of the corresponding Gross-Pitaevskii equation,
a symmetry-broken solution and a symmetry-preserving solution,
bifurcate from each other at a critical attraction strength \cite{Carr_2000,Peter_DW_2006}.
Thereafter, for stronger attractions,
the symmetry-broken mean-field solution becomes lower in energy
than the symmetry-preserving mean-field solution.

The implications of these properties are as follows.
For any number of particles $N$ and at constant interaction parameter $\Lambda$,
the energy per particle of the many-body fragmented ground state of the BEC is lower
than the energy per particle of the symmetry-broken mean-field solution \cite{CCI0_2004}
(see in this context \cite{Landman}).
At the infinite-particle limit,
the two energies per particle coincide \cite{Fragility_2008}.
It is the purpose of our fourth example to apply the concept
of many-body variance to such a situation.

\begin{figure}[!]
\begin{center}
\vglue -2.0 truecm
\hglue -1.25 truecm
\includegraphics[width=0.295\columnwidth,angle=-90]{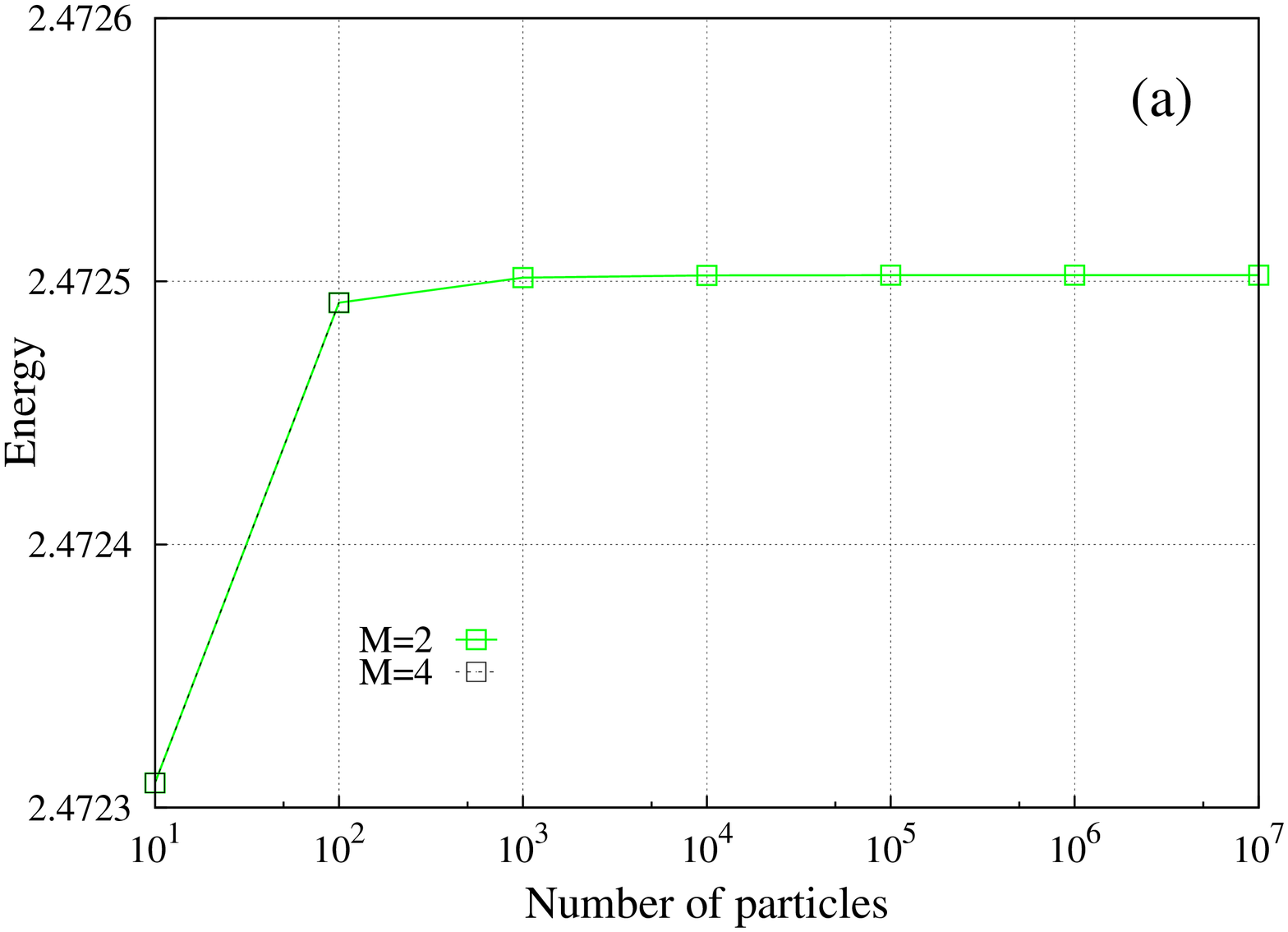}
\vglue 0.105 truecm
\hglue -1.0 truecm
\includegraphics[width=0.295\columnwidth,angle=-90]{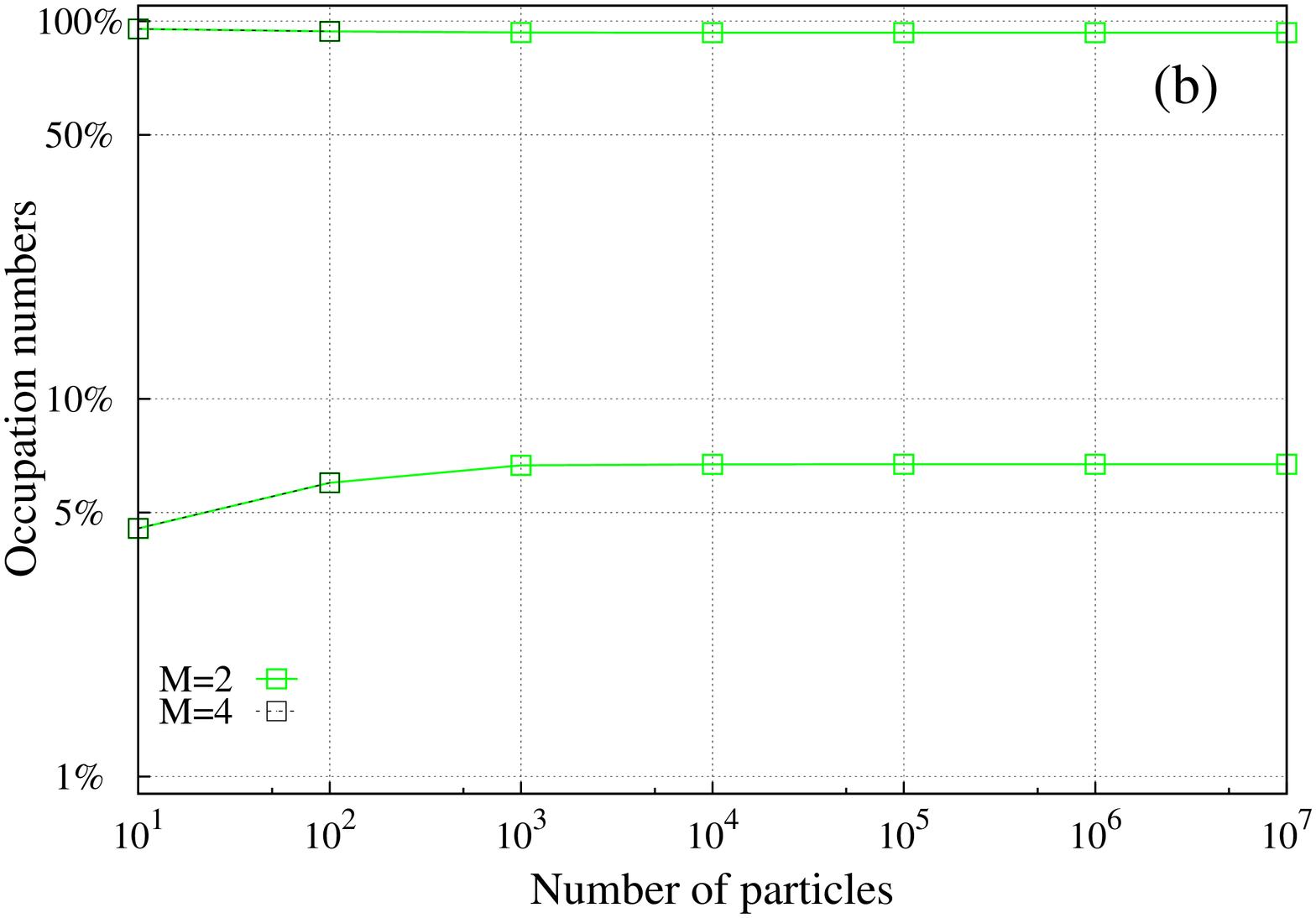}
\vglue 0.105 truecm
\hglue -1.0 truecm
\includegraphics[width=0.295\columnwidth,angle=-90]{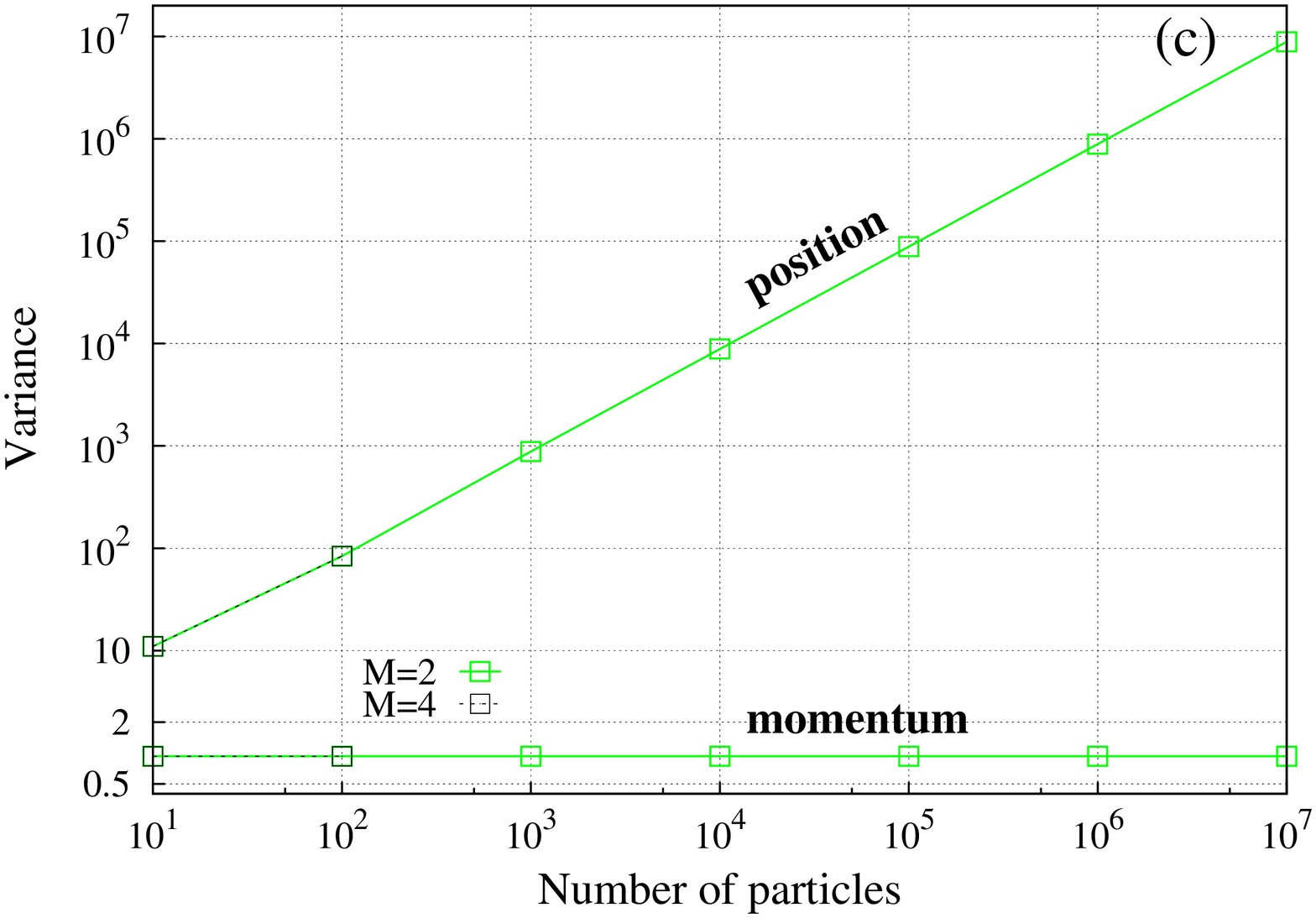}
\end{center}
\vglue 0.105 truecm
\caption{(Color online)
Ground-state properties of an attractive trapped BEC: Fragmentation.
Shown as a function of the number of bosons $N$ are:
(a) Energy per particle, $\frac{E}{N}$;
(b) Occupation numbers per particle $\frac{n_j}{N}$;
(c) Many-particle position variance per particle, $\frac{1}{N}\Delta^2_{\hat X}$, and
momentum variance per particle, $\frac{1}{N}\Delta^2_{\hat P_X}$.
The interaction parameter is $\Lambda=\lambda_0(N-1)=-0.018$,
corresponding to the weakest interaction strength in Fig.~\ref{f3}.
The confining potential is the double well $V(x)=0.05x^4+5.0e^{-0.5x^2}$.
Many-body results in all panels are computed at the $M=2$ level and marked by open green squares
(the connecting lines are to guide the eye only);
For demonstration of
convergence the results at the $M=4$ level for $N=10$ (see Fig.~\ref{f3} for $M>4$) and $N=100$ bosons
are marked by open black squares and seen to lie atop the $M=2$ data.
The results suggest that $\frac{E}{N}$ saturates [note the $y$ axis in panel (a)]
with increasing number of bosons $N$ and at constant $\Lambda$,
and that the BEC remains fragmented where the degree of fragmentation ($\frac{n_j}{N}$) saturates,
as expected from the best-mean-field approach for attractive BECs \cite{BMF}.
Furthermore, $\frac{1}{N}\Delta^2_{\hat X}$ increases (essentially) linearly
with the number of bosons and way beyond the system's physical size [note the $y$ axis in panel (c)],
whereas $\frac{1}{N}\Delta^2_{\hat P_X}$ saturates. 
See the text for further discussion.
The quantities shown are dimensionless.}
\label{f4}
\end{figure}

Coming back to the
system of attractive bosons in the double-well $V(x)=0.05x^4+5.0e^{-0.5x^2}$,
we consider the interaction parameter $\Lambda=\lambda_0(N-1)=-0.018$,
corresponding to the weakest interaction strength in Fig.~\ref{f3}.
We keep $\Lambda$ fixed and increase the number of particles $N$.
The number of particles is $N=10,100,\ldots,10\,000\,000$.
The number of orbitals is  $M=2,4,\ldots,12$ (including Fig.~\ref{f3}) for $N=10$,
$M=2,4$ for $N=100$,
and $M=2$ for $N\ge 1000$.
The results are collected in Fig.~\ref{f4}.
For the sake of analysis,
the BEC in the double well is solved also at the mean-field level ($M=1$) for the same interaction
parameter $\Lambda$.
Two energetically very close solutions,
symmetry-broken and symmetry-preserving, are found, see below.

In Fig.~\ref{f4}a,
the many-body energy per particle is depicted and seen to saturate as a function of $N$.
The energy per particle of the many-body solution
for $N=10\,000\,000$ bosons ($M=2$) is $\frac{E}{N}=2.472502328\underline{3}$.
It is lower than the energy per particle of the Gross-Pitaevskii symmetry-broken solution
only from the 10-th digit after the point on, $\frac{E}{N}=2.472502328\underline{4}$.
Indeed, as expected,
the many-body energy per particle 
approaches with increasing $N$ 
from below the energy per particle
of the Gross-Pitaevskii symmetry-broken solution.
Additionally,
the energy per particle of the symmetry-preserving mean-field solution,
$\frac{E}{N}=2.4725\underline{674266}$,
is very slightly higher than the above two energies,
indicating that we are just after the critical attraction
for the bifurcation of the two mean-field solutions.

In Fig.~\ref{f4}b we depict the two
largest occupation numbers per particles,
$\frac{n_1}{N}$ and $\frac{n_2}{N}$.
Similarly to the energy per particle in Fig.~\ref{f4}a,
they quickly saturate with $N$,
yet the system is fragmented.
We find that 
$\frac{n_1}{N}=93.29\%$ and $\frac{n_2}{N}=6.71\%$ ($M=2$) at the infinite-particle limit.
The results with $M=4$ orbitals for $N=10$ and $N=100$ particles
lie on top of the $M=2$ results,
indicating high convergence,
also see the inset in Fig.~\ref{f3}c.
Furthermore,
we find that
$\frac{n_{j\ge3}}{N}<3\times 10^{-8}$ for $N=10$ particles (see Fig.~\ref{f3})
and for ten-times more particles, $N=100$ ($M=4$),
that $\frac{n_{j\ge3}}{N}<3\times 10^{-9}$.
This decrease in the occupation numbers
strongly implies that 
the attractive BEC remains (only) two-fold
fragmented {\it en route} to the infinite-particle limit,
as expected from the best-mean-field approach for attractive BECs \cite{BMF}.

Finally, we present in Fig.~\ref{f4}c the many-body variances.
The many-body position variance per particle of the fragmented BEC, 
$\frac{1}{N}\Delta^2_{\hat X}$, is found to increase (essentially) linearly
with the number of bosons and way beyond the system's physical size.
This is in sharp contrast to attractive condensed BECs in the infinite-particle limit, see Fig.~\ref{f1}. 
On the other hand,
the many-body momentum variance per particle of the fragmented BEC,
$\frac{1}{N}\Delta^2_{\hat P_X}$, saturates, see Fig.~\ref{f4}c.
Explicitly,
we find numerically
$\frac{1}{N}\Delta^2_{\hat X}=0.88(5) \times N$ and
$\frac{1}{N}\Delta^2_{\hat P_X}=0.93$ in the infinite-particle limit ($M=2$).
The results with $M=4$ orbitals for $N=10$ and $N=100$ particles
lie on top of the $M=2$ results,
indicating high convergence,
also see the inset in Fig.~\ref{f3}d.
For reference,
the respective variances of the above-discussed
mean-field solutions are
$\frac{1}{N}\Delta^2_{\hat X}=\Delta_{\hat x, density}^2=3.7587$
and
$\frac{1}{N}\Delta^2_{\hat P_X}=\Delta_{\hat p_x, density}^2=0.9262$
for the symmetry-preserving Gross-Pitaevskii solution
and
$\frac{1}{N}\Delta^2_{\hat X}=\Delta_{\hat x, density}^2=2.8787$
and
$\frac{1}{N}\Delta^2_{\hat P_X}=\Delta_{\hat p_x, density}^2=0.9322$
for the symmetry-broken Gross-Pitaevskii solution.
It is remarkable that
fragmentation of less than $5\%$ and practically the same energy-per-particle
leads to such a macroscopic effect on the many-particle position variance.

All in all,
the uncertainty product $\Delta^2_{\hat X_{c.m.}}\Delta^2_{\hat P_{X_{c.m.}}} = 
\frac{1}{N}\Delta^2_{\hat X}\frac{1}{N}\Delta^2_{\hat P_X}$
of the ground state of a fragmented attractive BEC
diverges in the infinite-particle limit.
This is a good place to conclude the present investigation
of the intriguing physics of the many-body variance
of attractive trapped BECs.

\section{Conclusions}\label{Conclusions}

We have investigated on the many-body level 
the ground state and quench dynamics of structureless attractive bosons interacting by a 
finite-range (Gaussian) inter-particle interaction and in one-dimensional anharmonic trap potentials.
Thereby, we pay particular attention to the variance of the position and momentum 
many-particle operators. 
We have explicitly
(i) demonstrated how the shape of the anharmonic trap in combination with the inter-particle attraction affect the many-particle position and momentum variances,
(ii) investigated the behavior towards the infinite-particle limit of a trapped attractive BEC
by increasing the number of bosons up to $10^7$ bosons, its degree of condensation, and whether and how the many-particle variances computed at the many-body level differ from those computed at the mean-field level, 
both for the ground state and for an out-of-equilibrium quench scenario,
(iii) explored the many-particle variances of a fragmented
attractive BEC also towards its infinite-particle limit, and importantly,
(iv) proved numerically the high convergence of
the energy, density, depletion, occupation numbers, and of the position and momentum variances
for a broad class of scenarios, from condensation to fragmentation, of trapped attractive BECs.

\section*{Acknowledgements}

This paper is dedicated to Professor Dr. Wolfgang Domcke,
a dear friend and the first collaborator of one of us (LSC),
on the occasion of his 70th birthday.
This research was supported by the Israel Science Foundation (Grant No. 600/15).
We thank Sudip Haldar and Raphael Beinke for discussions. 
Computation time on the BwForCluster and the Cray XC40 
system Hazelhen at the High Performance Computing Center
Stuttgart (HLRS) is gratefully acknowledged.

\appendix

\section{Further computational details and convergence}\label{APP_Num}

\begin{figure}[!]
\begin{center}
\vglue -0.5 truecm
\hglue -1.0 truecm
\includegraphics[width=0.345\columnwidth,angle=-90]{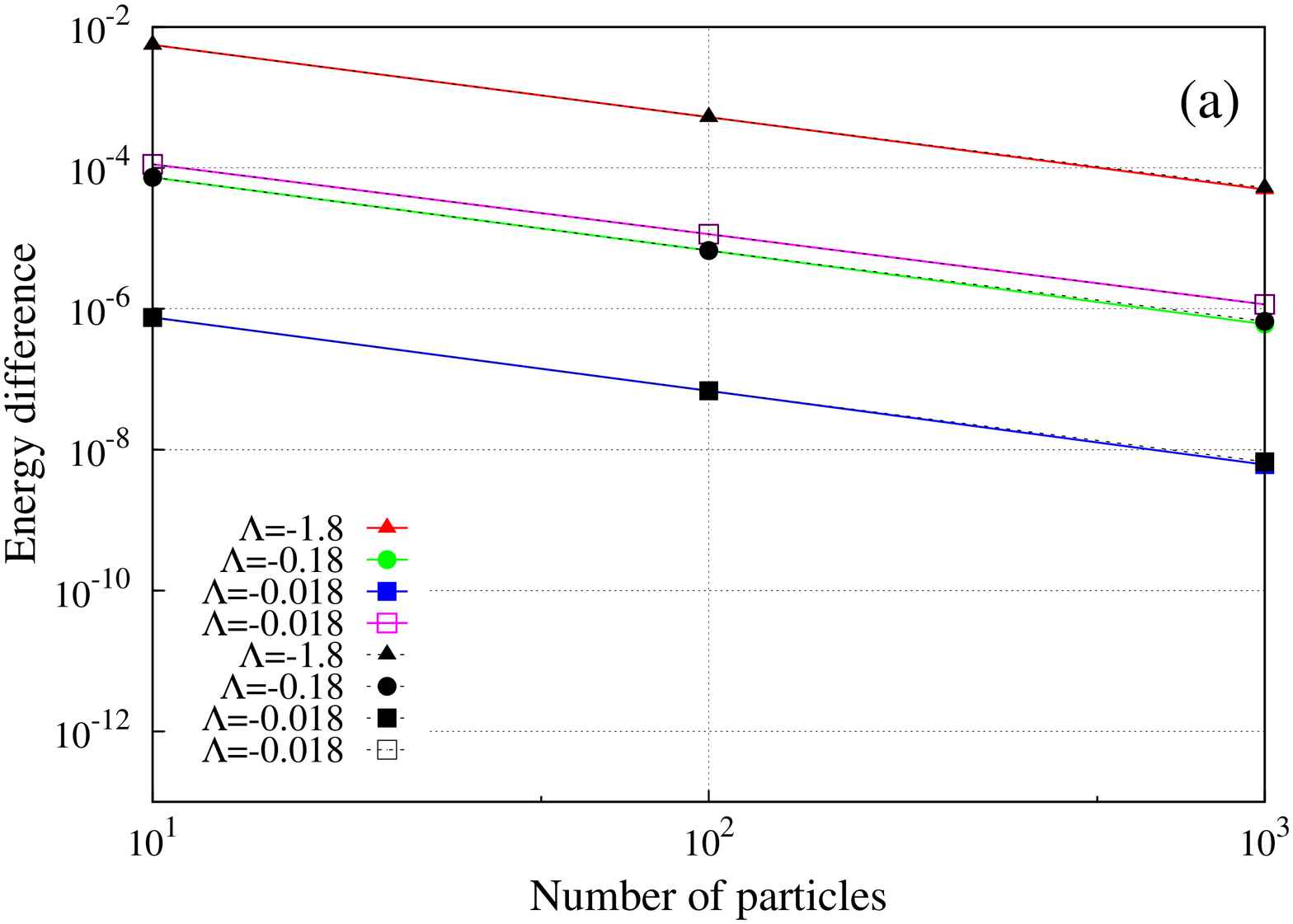}
\includegraphics[width=0.345\columnwidth,angle=-90]{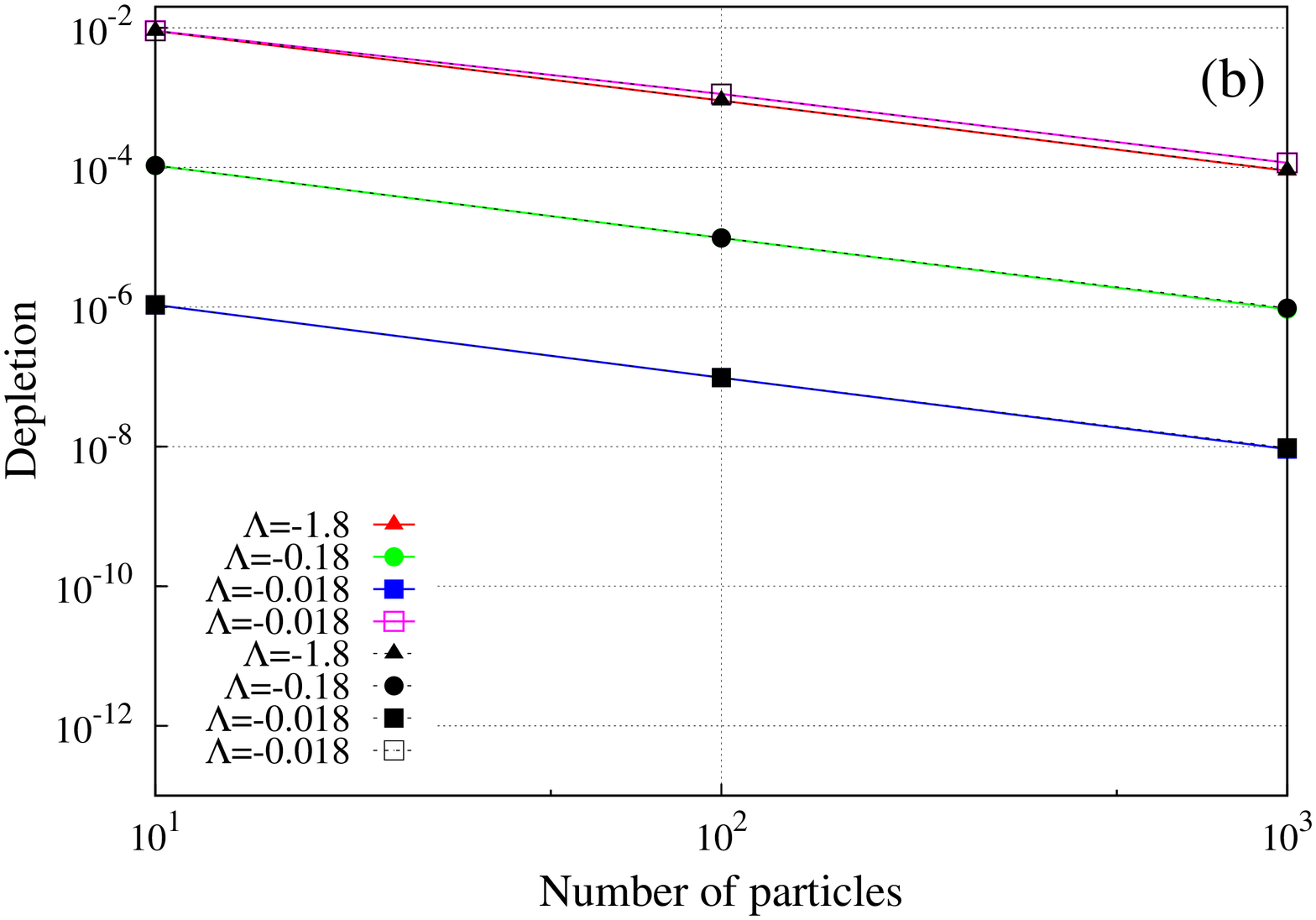}
\vglue 0.25 truecm
\hglue -1.0 truecm
\includegraphics[width=0.345\columnwidth,angle=-90]{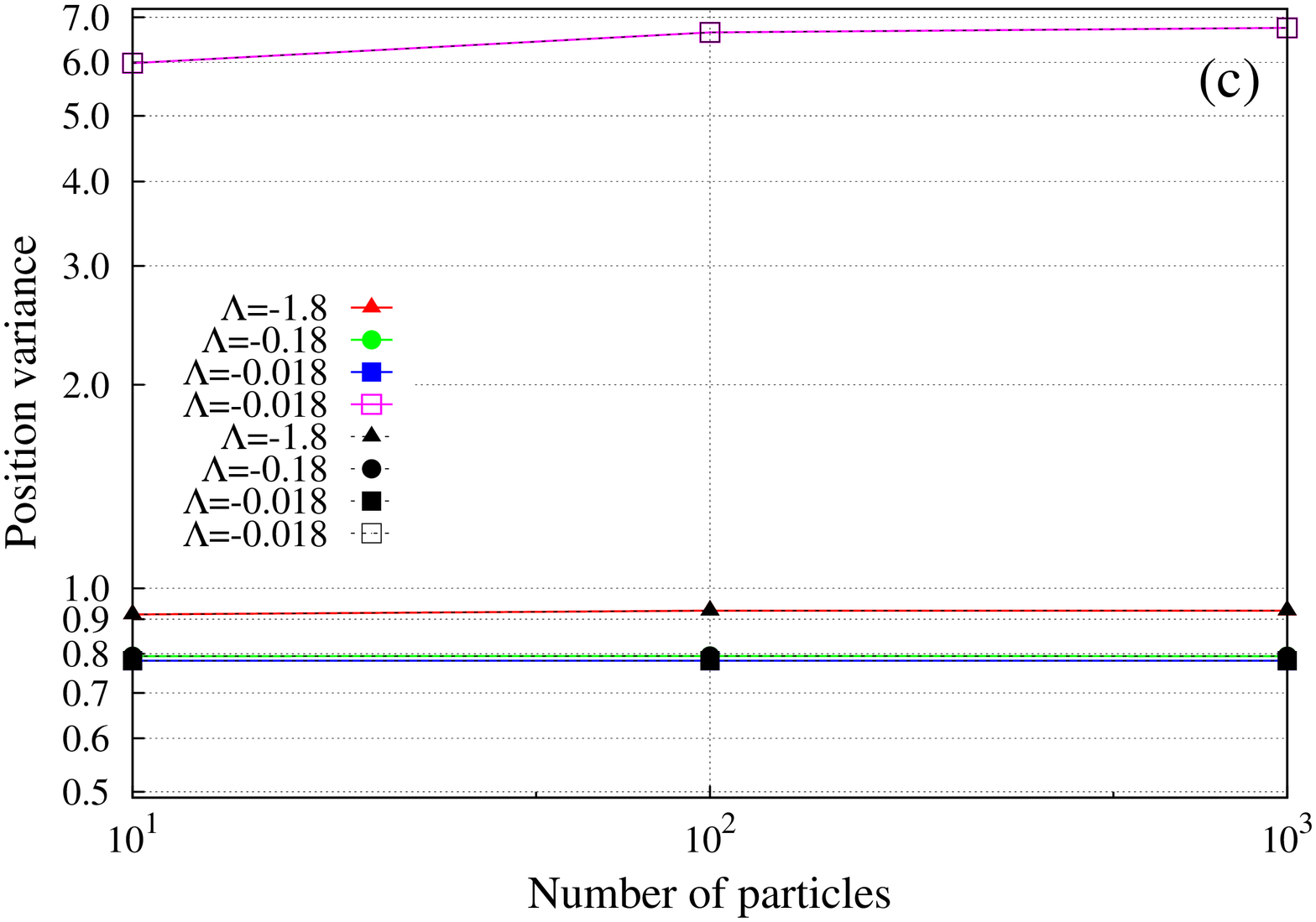}
\includegraphics[width=0.345\columnwidth,angle=-90]{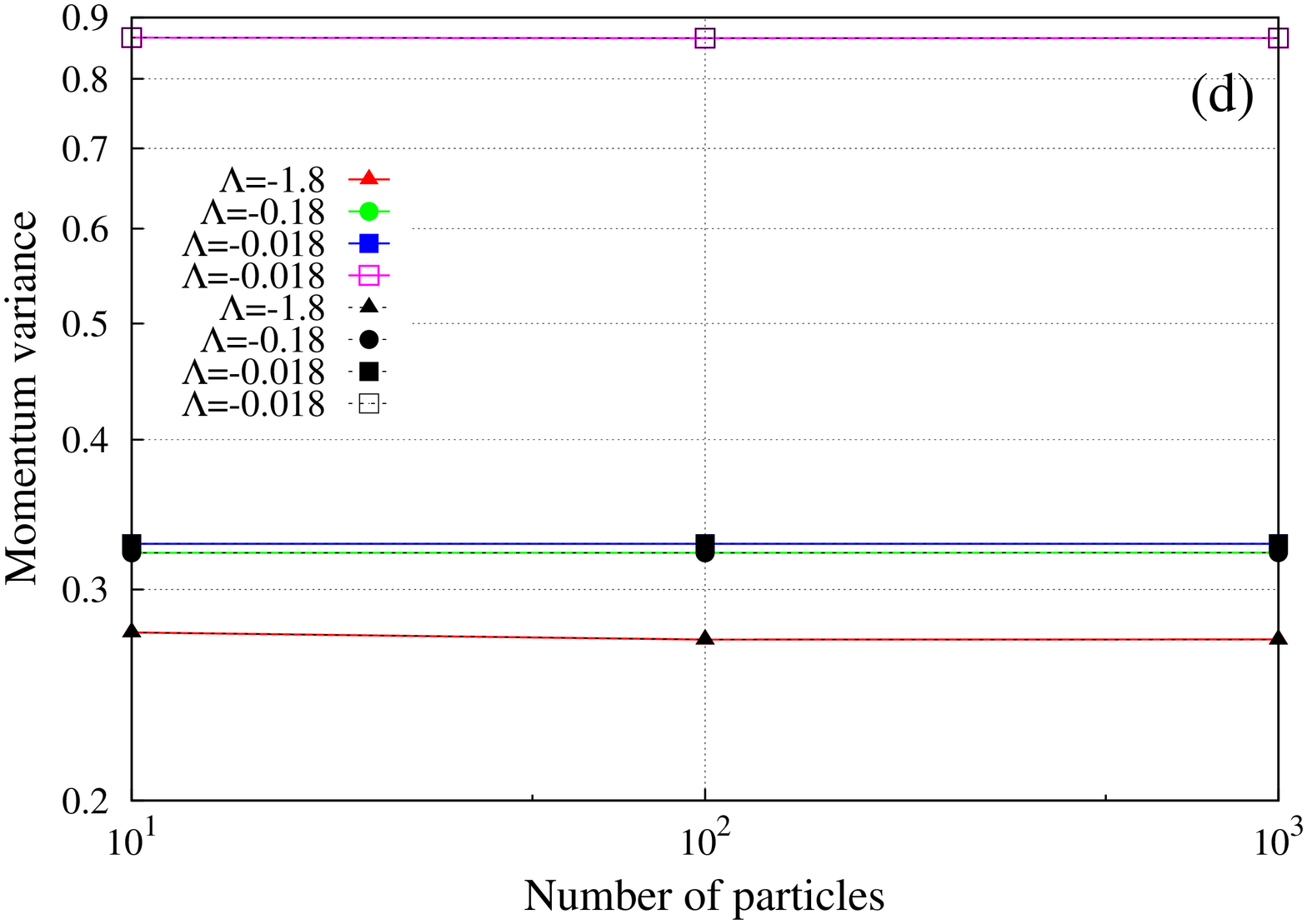}
\end{center}
\vglue 0.25 truecm
\caption{(Color online) 
Convergence of ground-state properties of an attractive trapped BEC: Condensation.
Shown for $N=10$, $N=100$, and $N=1000$ bosons 
as a function of $M$ self-consistent orbitals are:
(a) Difference between the mean-field and many-body energies per particle;
(b) Depleted fraction, $1-\frac{n_1}{N}$;
(c) Many-particle position variance per particle, $\frac{1}{N}\Delta^2_{\hat X}$;
(d) Many-particle momentum variance per particle, $\frac{1}{N}\Delta^2_{\hat P_X}$.
Interaction parameters are $\Lambda=\lambda_0(N-1)=-0.018$, $\Lambda=-0.18$, and $\Lambda=-1.8$.
The confining potentials are $V(x)=0.05x^4$ (filled symbols) and $V(x)=0.05x^4+4.5e^{-0.5x^2}$ (open squares).
The results for $N=10$ are with $M=5$,
for $N=100$ with $M=4$,
and for $N=1000$ with $M=2$,
all plotted by symbols in red, green, blue, and magenta colors.
The results with the respective successive number of orbitals, $M+1$, for all
data points
are plotted by the same symbol palette in black,
and seen to lie precisely atop, indicating high convergence of all properties.
All connecting lines, in colors and in black on top, are to guide the eye only.
See Fig.~\ref{f1} and the text for further discussion.
The quantities shown are dimensionless.}
\label{f5}
\end{figure}

\begin{figure}[!]
\begin{center}
\hglue -1.0 truecm
\includegraphics[width=0.2430\columnwidth,angle=-90]{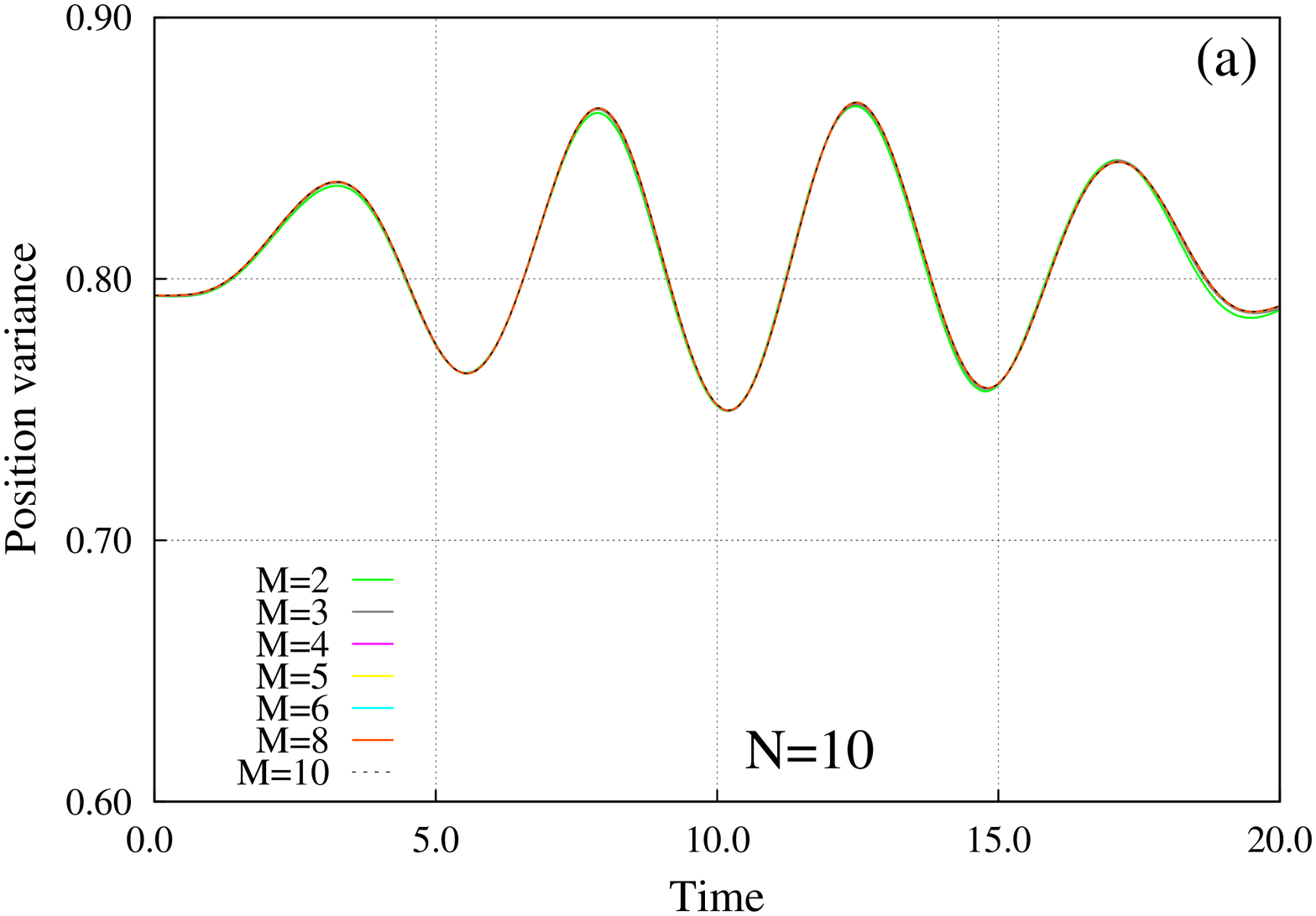}
\includegraphics[width=0.2430\columnwidth,angle=-90]{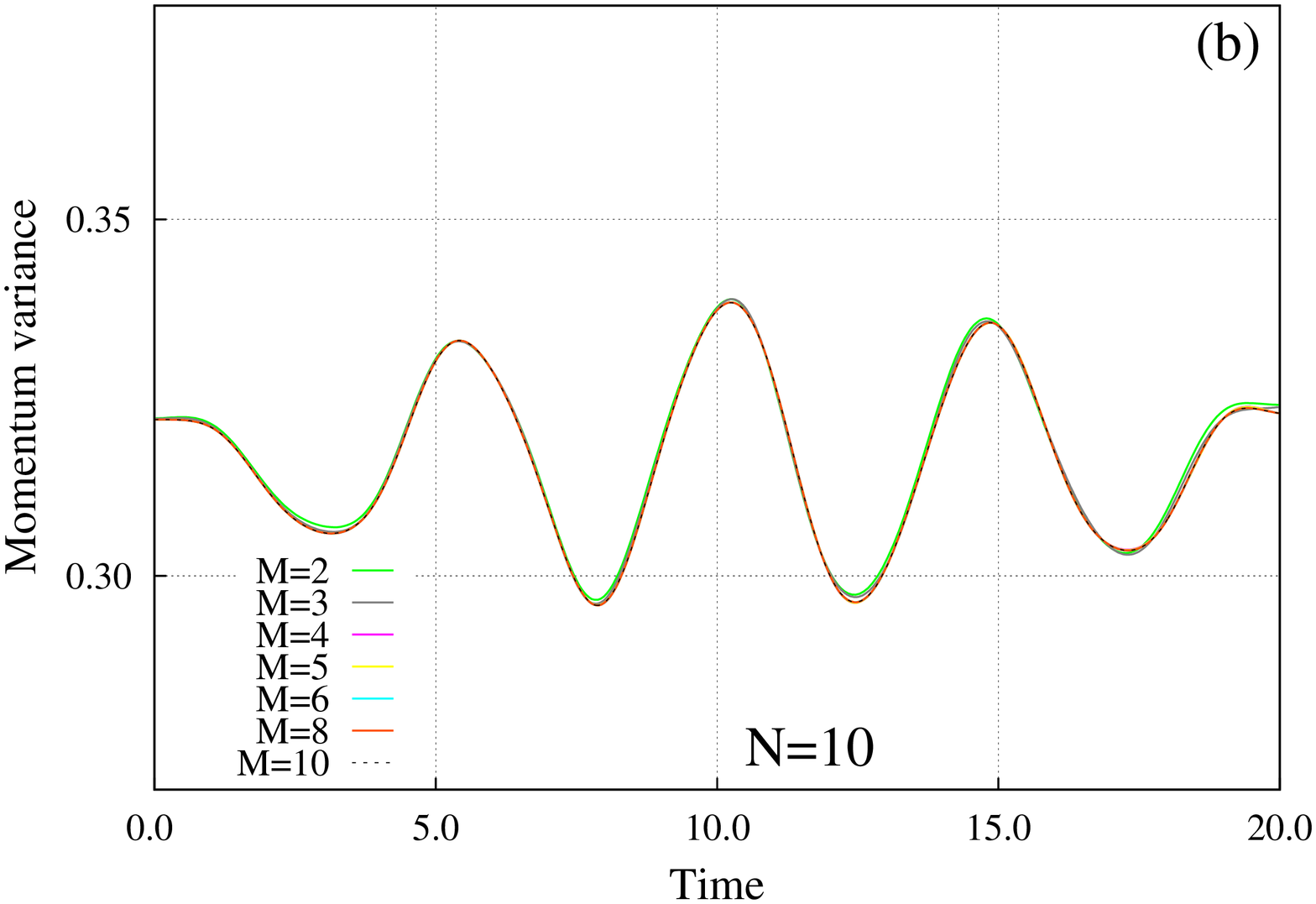}
\hglue 0.05 truecm
\includegraphics[width=0.2430\columnwidth,angle=-90]{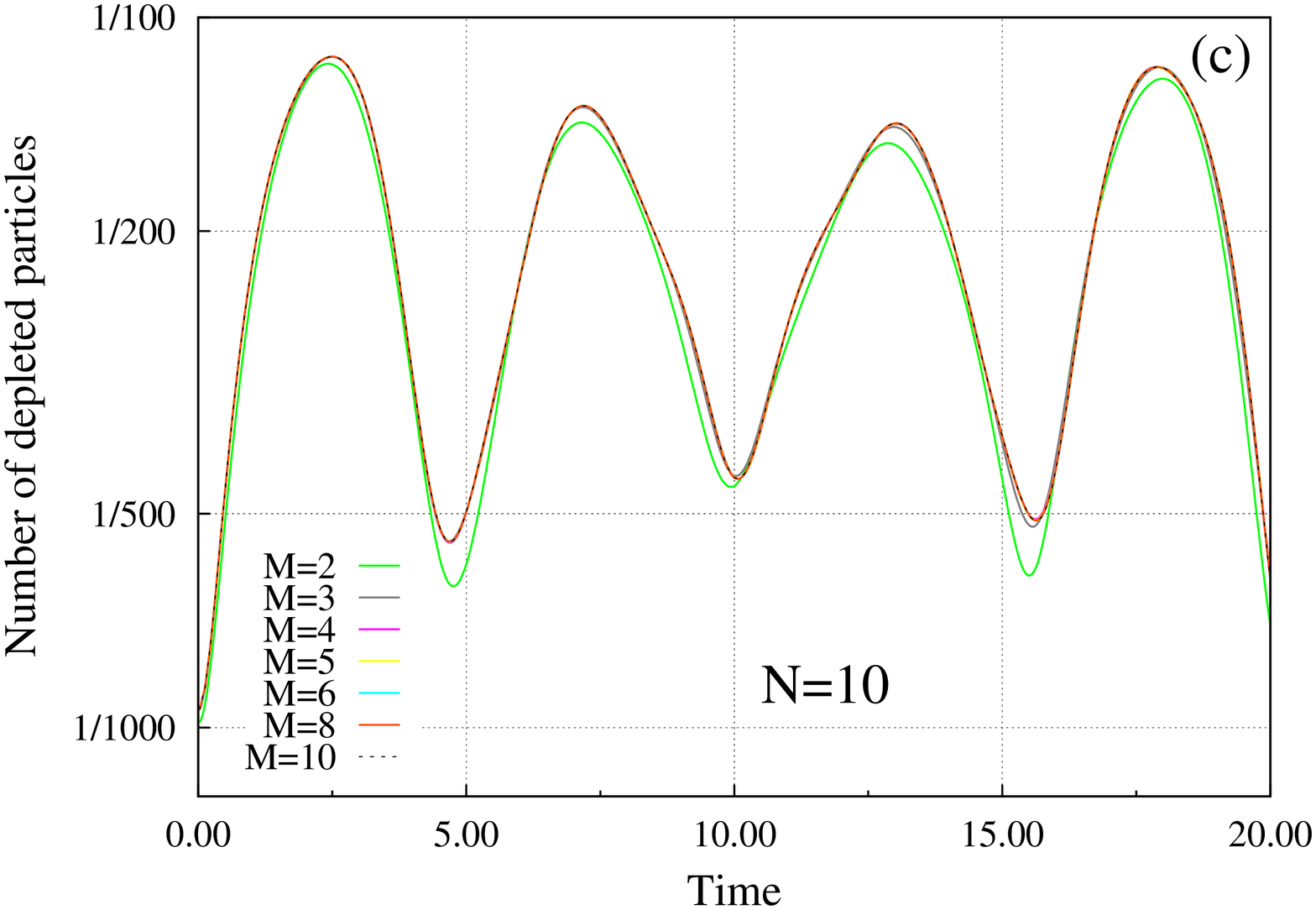}
\vglue 0.25 truecm
\hglue -1.0 truecm
\includegraphics[width=0.2430\columnwidth,angle=-90]{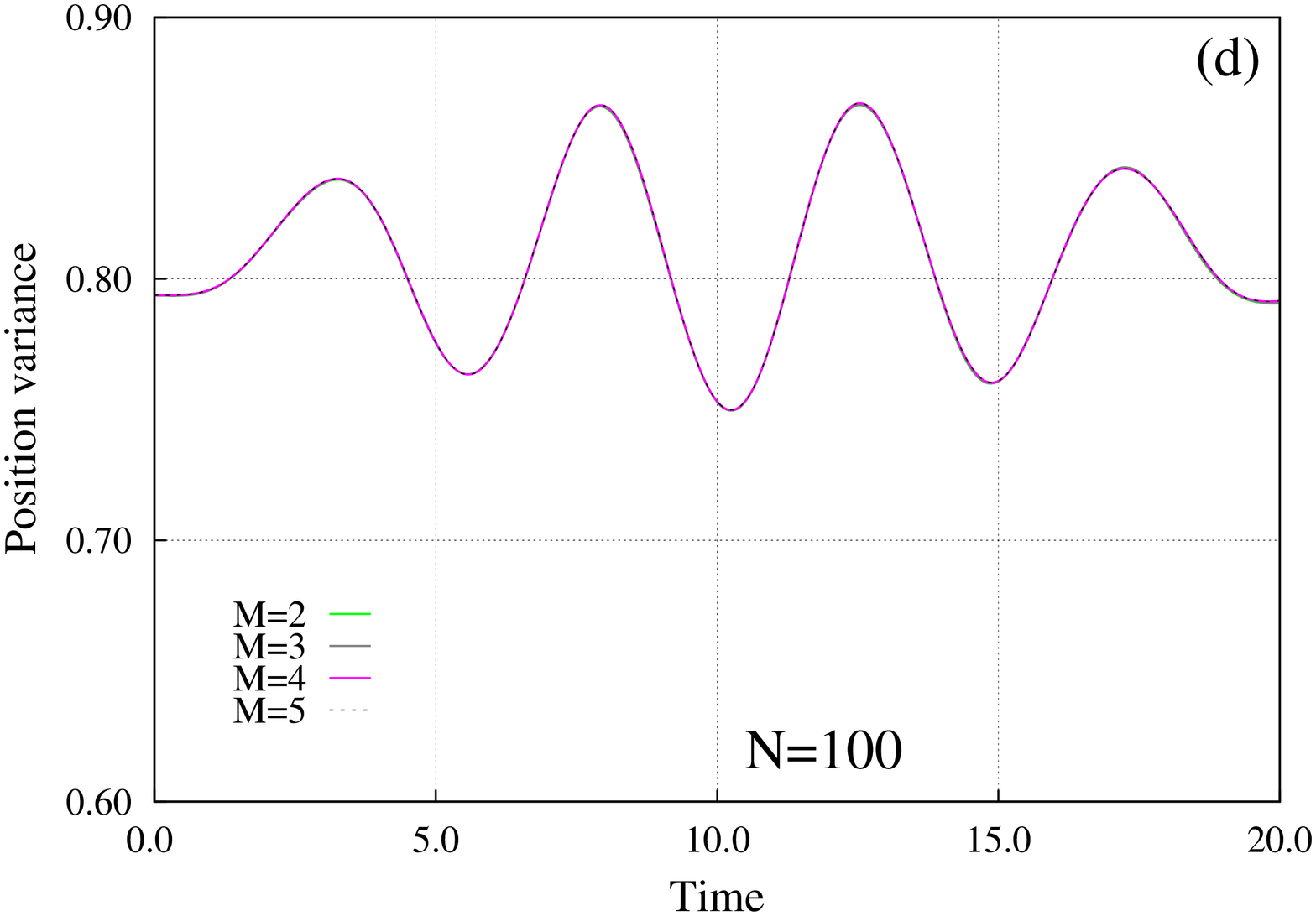}
\includegraphics[width=0.2430\columnwidth,angle=-90]{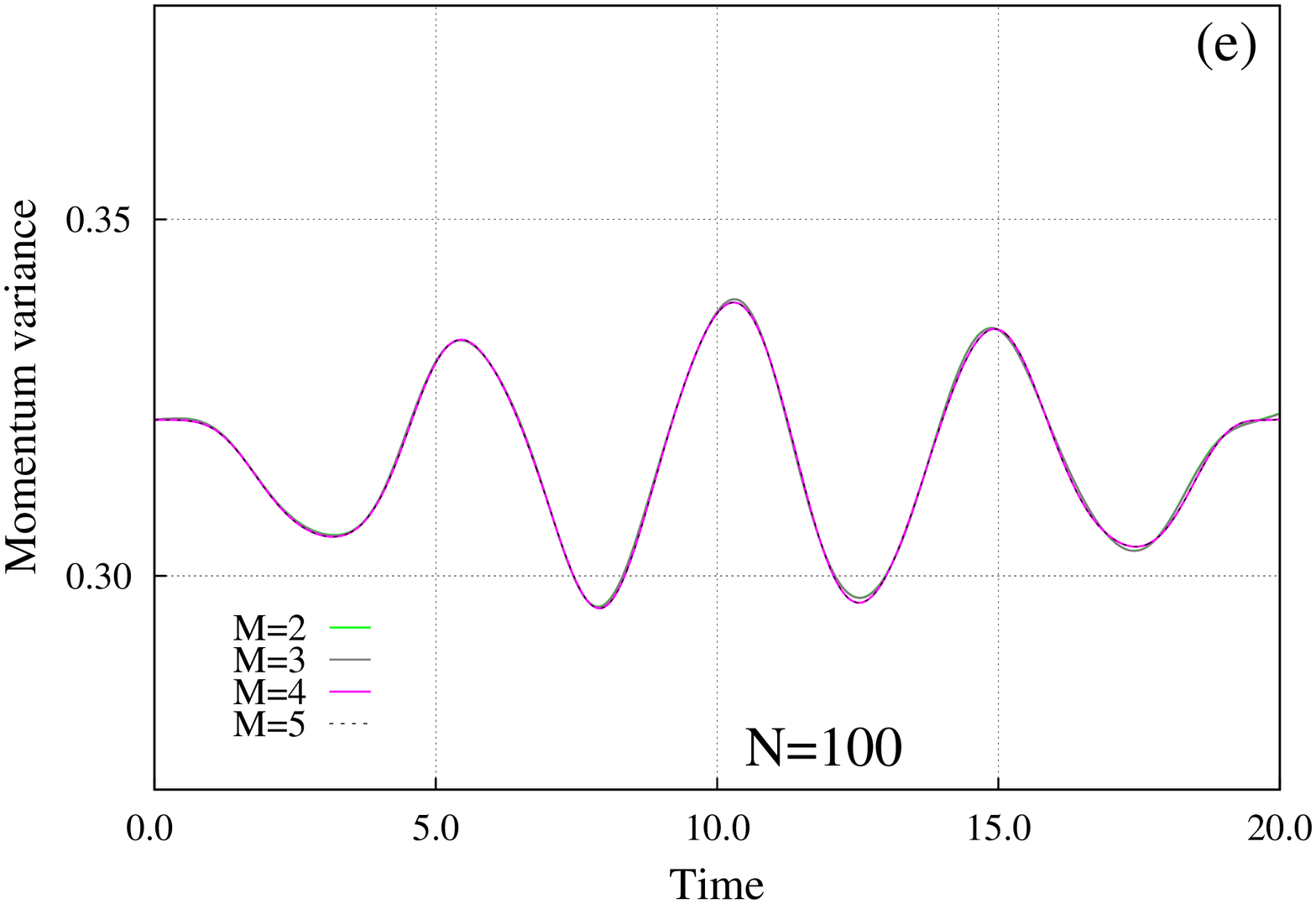}
\hglue 0.05 truecm
\includegraphics[width=0.2430\columnwidth,angle=-90]{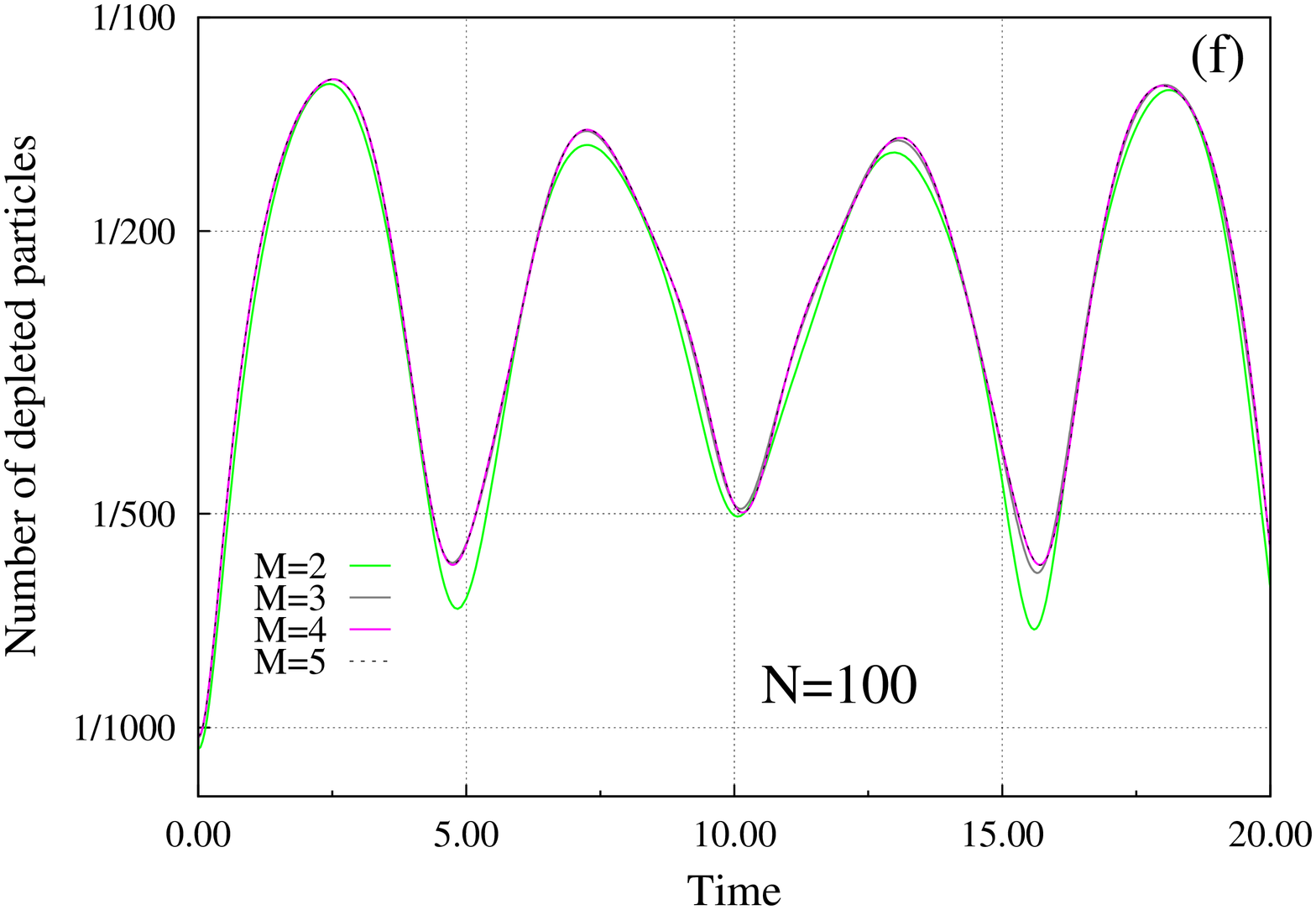}
\vglue 0.25 truecm
\hglue -1.0 truecm
\includegraphics[width=0.2430\columnwidth,angle=-90]{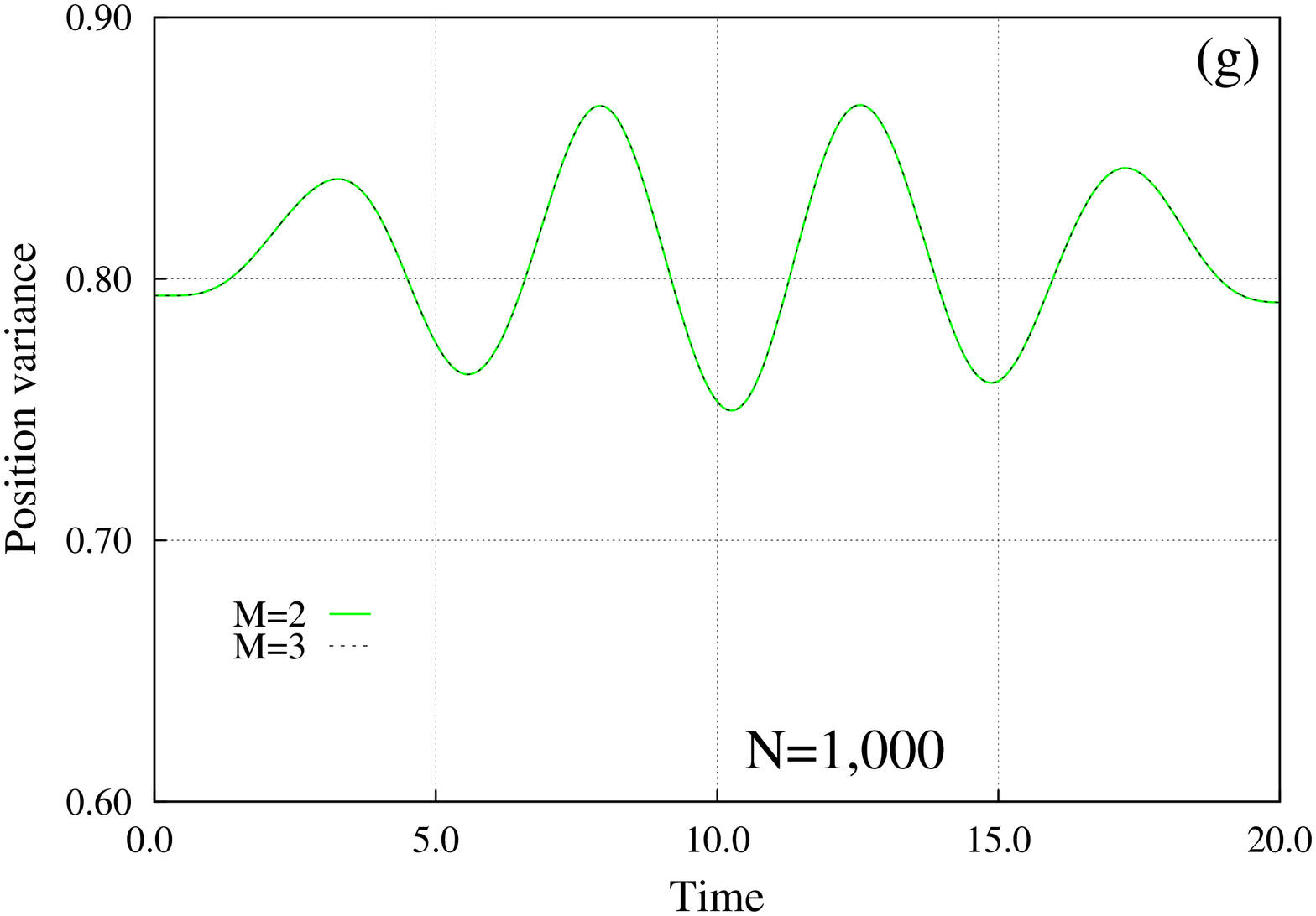}
\includegraphics[width=0.2430\columnwidth,angle=-90]{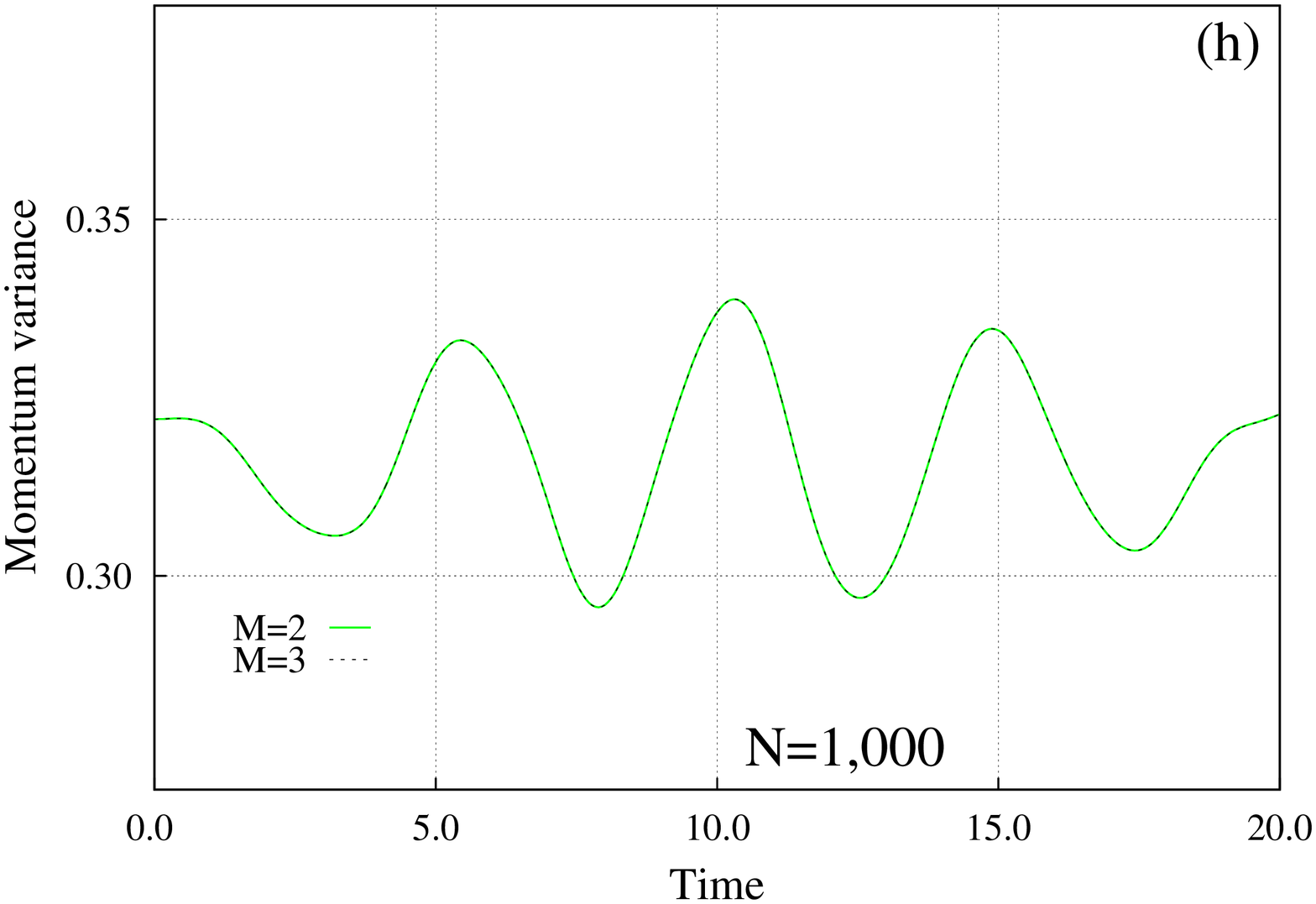}
\hglue 0.05 truecm
\includegraphics[width=0.2430\columnwidth,angle=-90]{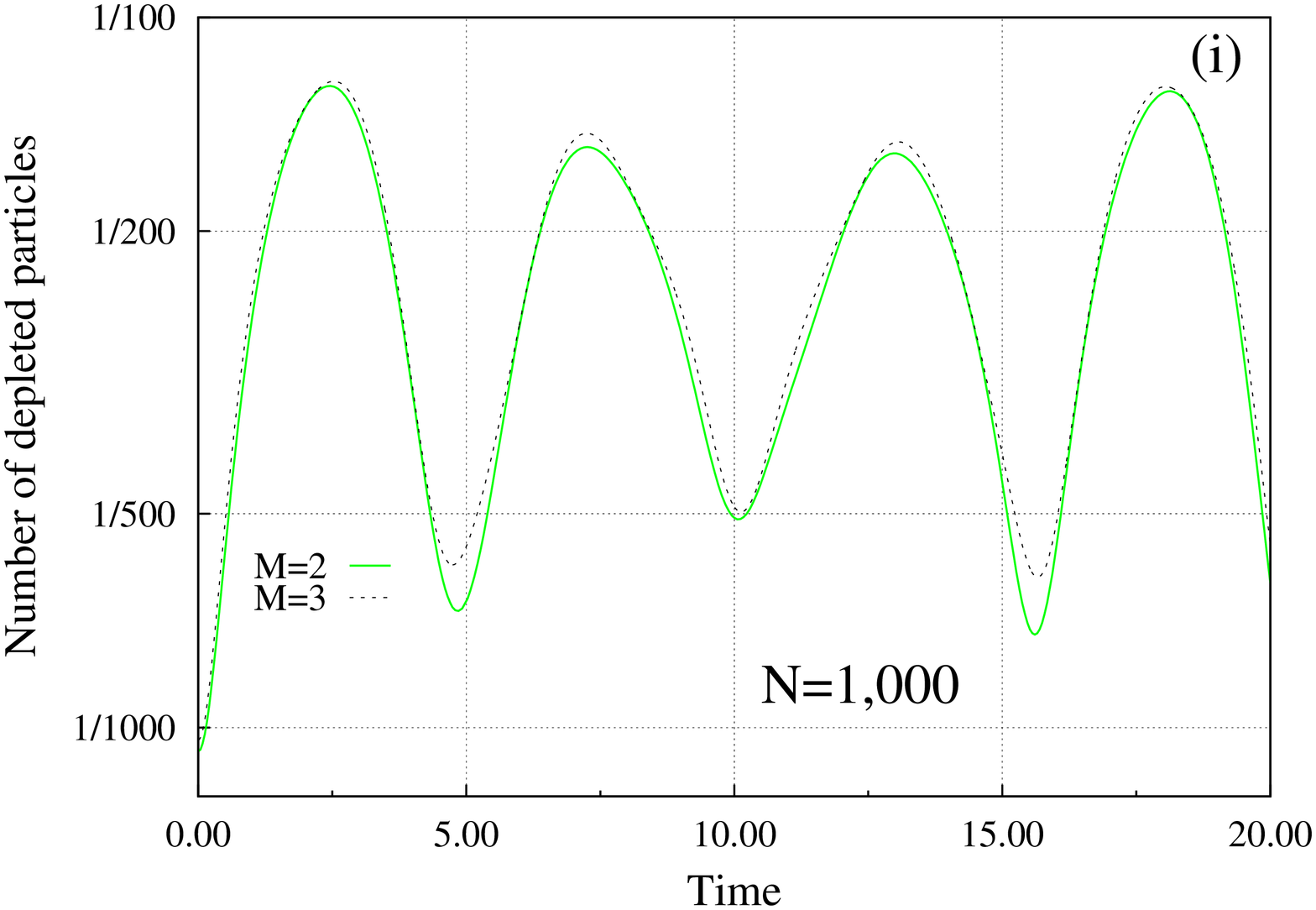}
\end{center}
\vglue 0.25 truecm
\caption{(Color online)
Convergence of the breathing dynamics of an attractive trapped BEC following an interaction quench.
Shown for $N=10$ [upper row, panels (a), (b), and (c)], 
$N=100$ [middle row, panels (d), (e), and (f)], 
and $N=1000$ [lower row, panels (g), (h), and (i)] bosons 
as a function of $M$ time-adaptive orbitals (smooth curves in color up to dashed curves in black on top) are:
Many-particle position variance per particle, $\frac{1}{N}\Delta^2_{\hat X}(t)$ [left column, panels (a), (d), and (g)];
Many-particle momentum
variance per particle, $\frac{1}{N}\Delta^2_{\hat P_X}(t)$ [middle column, panels (b), (e), and (h)];
Number of depleted particles, $N-n_1(t)$ [right column, panels (c), (f), and (i)].
The interaction parameter is quenched from $\Lambda=\lambda_0(N-1)=-0.18$
to $\Lambda=-0.36$ at $t=0$.
The confining potential is the single well $V(x)=0.05x^4$.
The time-dependent results for all quantities computed with $M=2$ time-adaptive
orbitals are already accurate as can be seen when comparing with the results of larger $M$.
The results obtained with $M=4$ ($M=3$) are already numerically well converged.
The many-particle position (momentum) variance initially increases (decreases),
in an opposite manner to the breathing of the density.
There is less than a $\frac{1}{100}$-th of a particle depleted.
See Fig.~\ref{f2} and the text for further discussion.
The quantities shown are dimensionless.}
\label{f6}
\end{figure}

The multiconfigurational time-dependent Hartree for bosons 
(MCTDHB) method \cite{MCTDHB1,MCTDHB2,Kaspar_The,book_nick,Axel_The,book_MCTDH,MCTDHB_LAB,package} 
is recruited in the present work to investigate the
ground-state and out-of-equilibrium properties
of attractive bosons in one-dimensional anharmonic traps
interacting by a finite-range (Gaussian) inter-particle interaction.
The maximal configurational space used are 
$\begin{pmatrix}1002 \cr 2\end{pmatrix}=501\,501$
for $N=1000$ bosons and $M=3$ orbitals,
$\begin{pmatrix}104 \cr 4\end{pmatrix}=4\,598\,126$ for $N=100$ bosons and $M=5$ orbitals,
and 
$\begin{pmatrix}21 \cr 11\end{pmatrix}=352\,716$ for $N=10$ bosons and $M=12$ orbitals.  
For the computations the many-body Hamiltonian is represented 
by $128$ exponential discrete-variable-representation grid points
(using a Fast-Fourier Transform routine)
in a box of size $[-8,8)$.
Convergence with respect to the grid size has been verified with $256$ grid points
and for the strongest inter-particle attraction strength.
Convergence of the energy, density, depletion, occupation numbers,
and the position and momentum variances
with increasing $M$ is numerically proved for the fragmented BECs in Figs.~\ref{f3}, \ref{f4}
of the main text,
and for the condensed ground-state and out-of-equilibrium dynamics in
Figs.~\ref{f5} and \ref{f6}.
In particular for the attractive trapped bosons,
convergence of the many-particle position and momentum variances 
for the ground state \cite{Variance}
and for the out-of-equilibrium breathing dynamics \cite{TD_Variance}
is nicely achieved and clearly demonstrated,  
also see in this context \cite{Brand_Cos}.

\end{document}